\documentclass[12pt]{iopart}

\usepackage{graphicx}
\usepackage{setstack} 
\usepackage{iopams}
\usepackage{mathrsfs}
\usepackage{amssymb}
\usepackage[T1]{fontenc}
\usepackage[british]{babel}

\begin{document}

\title{First-passage and first-hitting times of L\'evy flights and L\'evy walks}

\author{Vladimir V. Palyulin$^{\dagger}$, George Blackburn$^{\ddagger,
\pounds}$, Michael A. Lomholt$^{\flat}$, Nicholas W. Watkins$^{\pounds,\sharp,
\diamond,\oplus}$, Ralf Metzler$^\sharp$, Rainer Klages$^{\ddagger,\P,\S}$,
and Aleksei V. Chechkin$^{\sharp,\$}$}
\address{$\dagger$ Centre for Computational and Data-Intensive Science and
Engineering, Skolkovo Institute of Science and Technology, Nobelya Ulitsa
3, Moscow, 121205, Russia\\
$\ddagger$ Max Planck Institute for the Physics of Complex Systems,
N\"othnitzer Stra{\ss}e 38, D-01187 Dresden, Germany\\
$\pounds$ Centre for Fusion, Space and Astrophysics, University of Warwick,
Coventry, UK\\
$\flat$ MEMPHYS - Centre for Biomembrane Physics, Department of Physics,
Chemistry, and Pharmacy, University of Southern Denmark, 5230 Odense M, Denmark\\
$^{\sharp}$ Institute for Physics \& Astronomy, University of Potsdam,
D-14476 Potsdam-Golm, Germany\\
$\diamond$ Centre for the Analysis of Time Series, London School of Economics
and Political Sciences,  London, United Kingdom\\
$\oplus$ Faculty of Science, Technology, Engineering and Mathematics, Open
University, Milton Keynes, United Kingdom\\
$\P$ Institut f\"ur Theoretische Physik, Technische Universit\"at Berlin,
Hardenbergstra{\ss}e 36, 10623 Berlin, Germany\\
$\S$ Queen Mary University of London, School of Mathematical Sciences,
Mile End Road, London E1 4NS, United Kingdom\\
$\$$ Akhiezer Institute for Theoretical Physics National Science Centre
"Kharkov Institute of Physics and Technology", Kharkov 61108, Ukraine}
\ead{chechkin@uni-potsdam.de. Corresponding author: A. V. Chechkin}

\begin{abstract}
For both L{\'e}vy flight and L{\'e}vy walk search processes we analyse
the full distribution of first-passage and first-hitting (or first-arrival)
times. These are, respectively, the times when the particle moves across a
point at some given distance from its initial position for the first time,
or when it lands at a given point for the first time. For L{\'e}vy motions
with their propensity for long relocation events and thus the possibility to
jump across a given point in space without actually hitting it ("leapovers"),
these two definitions lead to significantly different results. We study
the first-passage and first-hitting time distributions as functions of the
L{\'e}vy stable index, highlighting the different behaviour for the cases
when the first absolute moment of the jump length distribution is finite or
infinite. In particular we examine the limits of short and long times. Our
results will find their application in the mathematical modelling of random
search processes as well as computer algorithms.\\[0.2cm]
Version: \today.
\end{abstract}

\section{Introduction}  

When a stochastic process $x(t)$ first reaches a given threshold value in
many scenarios follow-up events are triggered: shares are sold when their
value crosses a pre-set target amount, or chemical reactions occur when two
reactive particles encounter each other in space. The time $t$ at which this
triggering event first occurs, is either called the first-hitting (first-arrival)
or the first-passage time, as defined below
\cite{Redner,gleb,hughes}. While the physical analysis
of first-passage time problems has a long history, notably the seminal
works by Smoluchowski \cite{smoluchowski} as well as Collins and Kimball
\cite{collins}, even for the long-studied case of standard Brownian motion
\cite{BCKV18} significant progress has been achieved within the last decade
\cite{gleb,benichou,bresloff,holcman}. In particular, for finite domains
interesting results were unveiled for the mean and global mean first passage
times and their geometry-control \cite{benichou,Benichou10}. The often minute
particle concentrations inside biological cells and the related concept
of the few-encounter limit \cite{kolesov,otto,aljaz} motivated studies to
obtain the full probability density function (PDF) of first-passage times
in generic geometries, demonstrating strong defocusing (large spread of
first-passage times) and an intricate
interplay between geometry- and reaction-control \cite{aljaz,denis}.

\begin{figure}
\centering
\includegraphics[width=13.2cm]{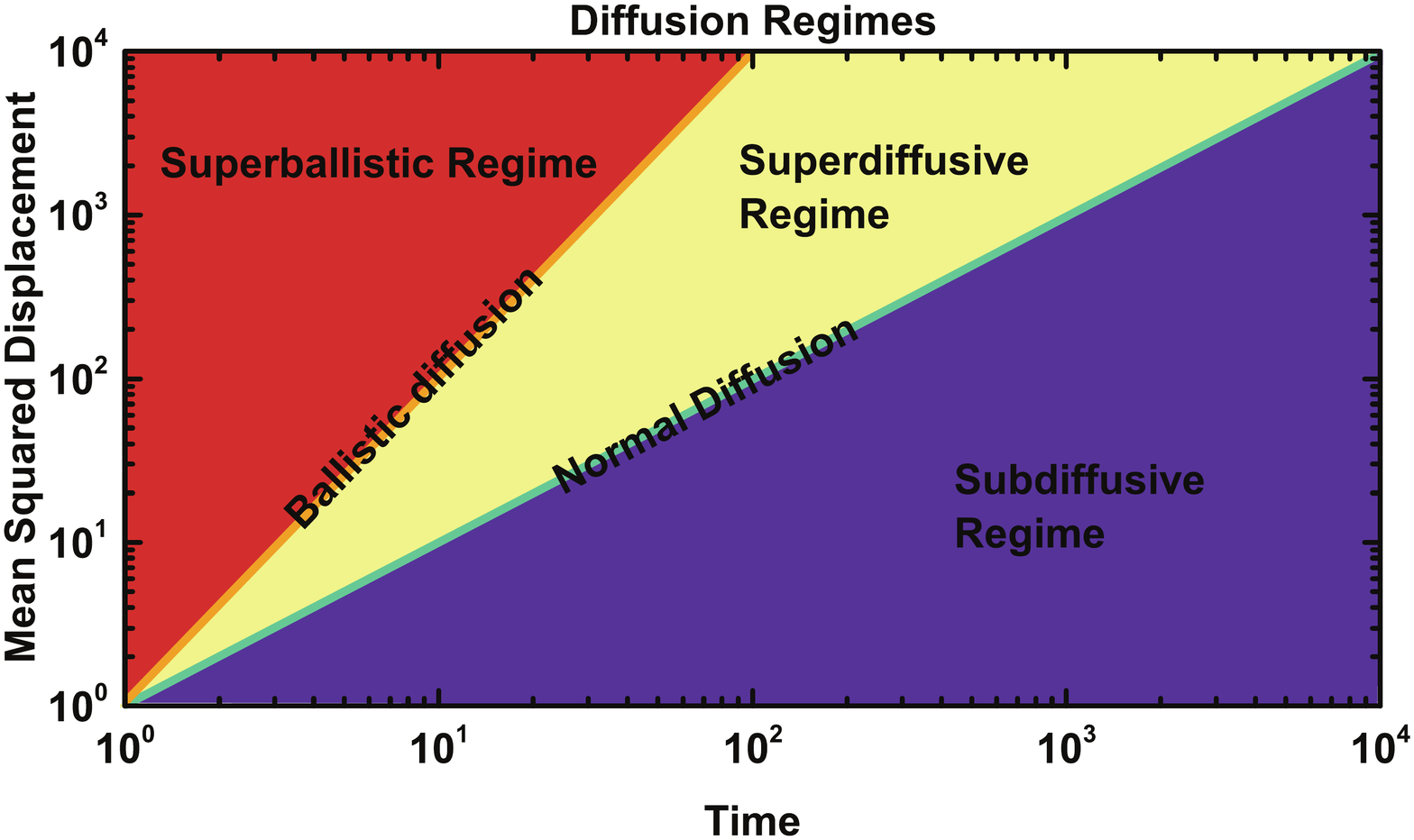}
\caption{Anomalous diffusion regimes as characterised by the power-law scaling
with time of the mean squared displacement (MSD). For normal diffusion the MSD
grows linearly with time. However, other types of "anomalous diffusion" are
widely observed, for which the MSD grows as $\langle x^2(t)\rangle\simeq t^\nu$,
where the anomalous diffusion exponent $\nu$ differs from unity. "Subdiffusion"
is slower than normal diffusion, corresponding to $\nu<1$. Faster diffusion splits
into the regimes of "superdiffusion" with $1 <\nu<2$, "ballistic motion" with
$\nu=2$, and "superballistic" with $\nu>2$.}
\label{LWdiagram}
\end{figure}

For non-Brownian stochastic processes additional complications in the
determination of first-passage and first-hitting times arise. Conceptually,
such \emph{anomalous diffusion processes\/} are distinguished according to the
value of the anomalous diffusion exponent $\nu$ in the long time limit of their
mean squared displacement (MSD) $\langle x^2(t)\rangle=\int x^2P(x,t)dx\simeq
K_{\nu} t^{\nu}$, defined as ensemble average of $x^2$ over the particle
PDF $P( x,t)$ \cite{report,MeKl04,KRS08,HoFr13,PCCP14,RevModPhys2015}:
the range $0<\nu<1$ corresponds to \emph{subdiffusion\/} (or dispersive
transport), while $1<\nu<2$ is referred to as \emph{superdiffusion\/} (or
enhanced transport). Beyond the regime $\nu=2$ of ballistic transport,
we encounter \emph{superballistic motion}. Sometimes the range $3<\nu<5$
is called \emph{hyperdiffusion\/} \cite{peter}. Examples for anomalous
diffusion processes include turbulent flows \cite{Rich26}, charge carrier
transport in amorphous semiconductors \cite{SM75}, human travel \cite{BHG06},
light waves in glassy material \cite{BBW08}, biological cell migration
\cite{DKPS08,HaBa12,Ariel,Andy}, or transport of submicron tracer particles
and fluorescently labelled molecules inside biological cells and
their membranes \cite{HoFr13,BGM12,Norregaard,BBABiomembranes2016}. Figure
\ref{LWdiagram} shows a schematic representation of the various regimes.

In anomalous diffusion the mentioned complications may arise due to long-ranged
correlations, for instance, in the increments of Mandelbrot's fractional Brownian
motion \cite{mandelbrot}, whose strongly non-Markovian character precludes the
application of standard analytic methods to determine the first-passage dynamics
\cite{molchan,jeon,igor}. Non-standard first-passage and first-hitting properties
also arise for \emph{L{\'e}vy flight (LF) and L{\'e}vy walk (LW)\/} processes,
that are in the focus of this study. LWs and LFs are among the most prominent
models for the description of superdiffusive processes \cite{report,MeKl04,KRS08,
RevModPhys2015,ch1,ch2,sp}. Both models represent special cases of continuous time
random walks \cite{hughes,klablushle}, in which relocation lengths are drawn from
a long-tailed L{\'e}vy stable distribution with diverging variance. The difference
between them is that LFs are Markovian processes in which jumps occur at typical
time intervals. Therefore, the
resulting LF process is characterised by a diverging MSD $\langle x^2(t)\rangle$
\cite{hughes,report,klablushle,hcf}. LWs, in contrast, include a spatiotemporal
coupling between jump lengths and waiting times, penalising long jumps with long
waiting times, effectively introducing a finite velocity \cite{ShKlafterWong}.
Jump lengths and waiting times may be coupled linearly, such that the resulting
LW moves in a given direction with a constant speed until velocity reversal after
a given waiting time, the velocity model \cite{ZuKl93a}, or the space-time
coupling may have a power-law type \cite{klablushle}. The velocity may also be
considered to change from one step to another \cite{2012,RadonsPRL2018}.
Interestingly, for certain parameters even L{\'e}vy walks
have an infinite variance, namely, when there is a distribution of velocities
associated with different path lengths \cite{RadonsPRL2018}.
LFs and
LWs are non-ergodic in the sense that long time and ensemble averages of physical
observables are different \cite{PCCP14,RevModPhys2015,RadonsPRL2018,daniela,daniela1,
aljaz1,aki}. 
The linear response behaviour and time-averaged Einstein relation of LWs have been
studied, as well \cite{daniela2,aljaz2}. We note that LFs and LWs have also been
formulated in heterogeneous environments \cite{epl,srokowski}.

\begin{figure}
\centering
\includegraphics[width=10cm]{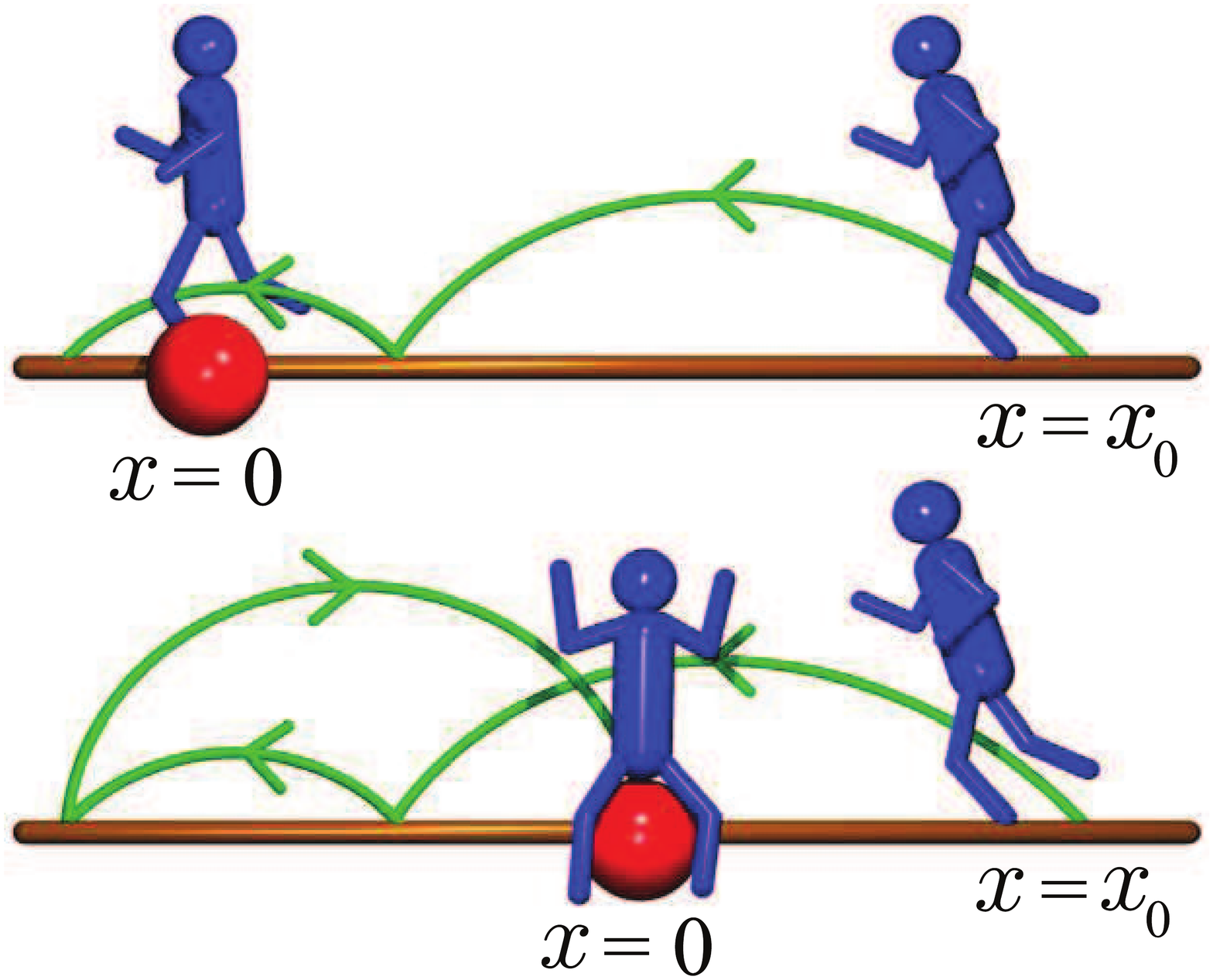}
\caption{Illustration of the difference between an event of first-passage (top),
when the walker crosses a target at $x=0$ in a single jump event and overshoots
it by the leapover length \cite{leapover2007}, and an event of first-hitting
(first-arrival, bottom) at the target. The initial position of the walker is
$x=x_0$.}
\label{PassageVSArrival}
\end{figure}

While for Brownian motion the events of first-passage and first-hitting (or
first-arrival) are identical because space is being explored continuously
\cite{Gardiner}, the possibility of long, non-local jumps lead to "leapovers"
\cite{koren,leapover2007}, single jumps in which a given point is overshot by some
leapover length, as illustrated in figure \ref{PassageVSArrival}: for random
walk processes with diverging variance of the jump length PDF, the event of
first-passage becomes fundamentally different from that of first-hitting, and
it is intuitively clear that first-hitting a target is harder (less likely) than
the first-passage.

LFs and LWs have been studied for considerable time, however, the systematic
comparison of some of their first passage properties started only recently
\cite{BartokEli2017}. We here systematically investigate the first-passage and
first-hitting properties of LFs and LWs in one dimension by drawing generic
conclusions on their differences and similarities. This is an important first
step in the assessment of these two fundamental random search processes.
Further applications of non-local, L{\'e}vy-type search are found
in computer algorithms such as simulated annealing \cite{ilya}.
In higher
dimensions the comparison is more complicated due to various possible definitions
of LWs and LFs \cite{Marek2012,EliMult}. In particular, so far studies of LFs and
LWs mostly focused on the long-time features. However, in all search processes the
short-time properties do contribute to the efficiency of the search \cite{JPA16},
and in the light of the aforementioned scenarios of the few-encounter limit become
even more relevant. We therefore pay considerable attention to the characteristics
of the short-time first-passage and first-hitting properties of LFs and LWs.

Our paper is organised as follows. In section 2 we briefly review the r{\^o}le of
LFs and LWs in random search processes, followed by setting the first-passage and
first-hitting scenarios in section 3. Section 4 reports the first-passage properties
of LFs and LWs, the first-hitting properties of both processes are then investigated
in section 5. A summary and discussion is provided in section 6, and details of the
mathematical derivations are deferred to the appendices.

\section{The role of L{\'e}vy flights and walks in random target search}

The term LF was coined by Beno{\^i}t Mandelbrot in honour of his advisor, the
French mathematician Paul L{\'e}vy, at {\'E}cole Polytechnique in Paris. In
his famed treatise on the fractality of nature \cite{mandelbrot2}, Mandelbrot
studied random walk processes with scale-free jump length distributions, leading
to fractal trajectories of clusters of local motion interspersed with long
relocations, on all scales.

As mentioned above LFs are Markovian and possess a diverging MSD. In that sense
they are in most cases "unphysical", as they appear to possess an infinite speed.
While spatiotemporally coupled LWs provide a "non-pathological" description with
finite speed, LFs do have their justification in the following senses: first, when
the process involves long-tailed jump lengths in some "chemical" co-ordinate but
local jumps in the physical, embedding space, the argument about a diverging MSD
does not hold. An example is the random search of proteins on a DNA chain that is
represented as a fast-folding chain in three dimensions \cite{Lomholt2005}. Second,
when we solely speak about the spatial trajectory described by the searcher yet
are oblivious of the corresponding time trace, it is legitimate to speak of LFs.
Third, the observed motion may be considered scale-free only within a limited
range of relocation lengths, beyond which cutoffs may exist \cite{mantegna,
honkonen,chech}. Fourth, we note that only very rarely have experimental foraging
studies really tested for LWs \cite{RevModPhys2015,Kla16}. Fifth, in some physical
systems the measured data indicate a divergence of the kinetic energy \cite{walter}.
Finally, LWs represent a much harder mathematical problem than LFs while the LF
assumption may already provide valuable insight into the system. For these reasons
we study LFs and LWs in parallel, and we point out their commonalities and
differences.

Interest in both LF and LW models in physics arose in the context of simple
one-dimensional deterministic maps that generate superdiffusive motion encountered
within the context of phase diffusion in Josephson junctions \cite{GeTo84,GNZ85}.
It was then shown that diffusion in these maps could be understood in terms of LWs
\cite{ZuKl93a,GeTo84,GNZ85,ShlKl85}. Further motivation to investigate LFs and LWs
was sparked by a remark of Michael Shlesinger and Joseph Klafter, that scale-free
motion may present a more efficient means for exploring space in one and two
dimensions as compared to Brownian motion \cite{ShleKlaft1986}. The argument goes
that while Brownian dynamics features repeated returns to already explored regions
causing oversampling \cite{hughes,Lomholt2005}, scale-free LFs and LWs, in
contrast, avoid these returns and, hence, yield a more efficient search
strategy.\footnote{Note that many search processes indeed run off in effectively
two dimensions (land-bound animals, birds or fish whose lateral motion has a much
wider span that the vertical motion) or even one dimension (proteins searching a
DNA, animals foraging in the border region between forests and grassland).}

The high efficiency of L{\'e}vy search became famous when experimental data of
the relocation distances of soaring albatross birds were reported to display
long, power-law tails \cite{Vis96}. While this result was discussed
controversially in the literature \cite{Edw07,albatross2012}, it prompted the
proliferation of the {\em L{\'e}vy flight foraging hypothesis}. Roughly speaking,
this hypothesis predicts that search processes with L{\'e}vy stable relocation
length distributions provide an optimal search strategy by minimising the random
search times under certain conditions, particularly for a low density of targets
\cite{Kla16,Vis99,VLRS11}. Although later many of the experimental studies testing
the L{\'e}vy flight foraging hypothesis were found to contain experimental
or methodological errors \cite{RevModPhys2015,Kla16,Intermittent,Pyke15}, there
exists ample evidence that many animals in fact do exhibit scale-free movements
over a few orders of magnitude \cite{VLRS11}, or have at least a search component
that is scale-free \cite{albatross2012}. From a theoretical point of view, however,
it was shown that even in the case of single-mode search LFs may not always optimise
a suitably defined search efficiency \cite{PNAS14,JStatMech14}. For example, if a
target is located in the close vicinity of a starting point or the target lies
"windward" (in the direction of an external bias) from the searcher's starting
position, Brownian motion may outperform the search by LFs. Furthermore, depending
on the precise biological and ecological conditions, often more complex search
patterns, for instance, multimodal or intermittent search strategies, are superior
to LFs and LWs \cite{ShKlafterWong,JPA16,Intermittent,EPJB17,benichou2006,gleb1,
LomholtPNAS2008,Lukasz}. Generally, most of the intermittent and multimodal search
strategies can be described as a combination of a local exploration mode and a
scale-free (long relocation) mode \cite{Intermittent,Benh99}, where the frequent,
long, relocations are described by L{\'e}vy motions \cite{JPA16,EPJB17}.

Interestingly, there exists a direct connection between the biological strategy
of random search for targets and the mathematical problem of first-passage and
first-hitting. In foraging theory one distinguishes between cruise and saltatory
foragers. While in cruise search the forager looks out for targets during the
movements, the forager is blind for its targets when moving for saltatory search,
for which the searcher needs to land exactly on (or very close to) the target to
actually detect it \cite{PNAS14,JPP09}. A cruise forager searching for a single
target, however, is modelled mathematically as a first-passage problem while
saltatory search for a single target corresponds to a first-hitting problem.
Solving these mathematical problems thus sheds light, at least in more simple,
abstract settings, on specific aspects of biological foraging.

As a  general rule first-hitting problems appear in most target search problems
while first-passage relates to the crossing of thresholds. The latter problem has
been studied for many decades in the context of Markovian processes (see, for
instance, \cite{Siegert}) and much initial work related to the Gaussian case.
Classic applications of first-passage include vibration \cite{He}, sea states,
and earthquakes \cite{ShinozukaWu}. The first-passage problem also appears in
calculating the time to activation in cases like Gerstein's and Mandelbrot's model
of neurons \cite{GersteinMandelbrot}, and resonant activation \cite{DybGod2009}.
Further motivation comes from applications of the area swept out by Brownian
motion up to its first passage, where results can be heuristically motivated from
first-passage theory scaling results \cite{KearneyMajumdar2005}. Recent interest
has included transitions from the bull-to-the-bear market in finance, extreme value
statistics of temperature records \cite{eichner2006}, or the diffusion of atoms
in a one-dimensional periodic potential \cite{kindermann2017} as well as problems
of escaping an enemy territory. The existence of heavy-tailed log-price fluctuations
in finance, modelled using $\alpha$-stable processes in \cite{mandelbrot1} has
indeed been a strong initial motivation for the LF first passage problem.

\section{Setup of the system: determining first-passage and first-hitting times}

The problem of first-passage consists in the determination of boundary crossing
dynamics of a searcher. In the general case first-passage properties can be studied
for any kind of domains \cite{Redner,gleb,Benichou10,aljaz,denis}. In this paper we
consider the classical 1D setting. We are interested in the event when a searcher
crosses the origin for the first time after initially being released at position
$x_0>0$ (figure \ref{PassageVSArrival}, upper part). For LFs the first-passage time
corresponds to the moment in time when the searcher first hits a coordinate on the
negative semi-axis. For LWs the first-passage time is defined as the time needed
for a particle to reach the origin, that is, only the fraction of the last relocation
event (from the arrival point of the previous relocation to crossing the origin) is
included in the computation.

A first-hitting event occurs when the end point of a jump arrives exactly at the
origin. In physics literature the first-hitting is often also called the first
arrival \cite{PNAS14,JPA2003}. Here we will use the term hitting throughout the
text to avoid possible ambiguities. Similar to the first-passage the event
of first-hitting may be defined for a domain of any shape. Here we concentrate
on point-like targets. Thus, in the case of LFs the first-hitting corresponds to
an exact landing at the target coordinate. For LWs in our 1D scenario an event
of first-hitting occurs when the last relocation ends at the target, that is,
in the LW case this corresponds to a searcher who cannot identify the target
while crossing it during the relocation event.

Our results are either derived from the fractional Fokker-Planck equation
\cite{report,hcf,epl} or are obtained from simulations of the discrete Langevin
equation
\begin{equation}
x_{n+1}-x_n=\left(K_\alpha \delta t\right)^{1/\alpha}\xi_{\alpha}(n)
\label{LEfinal}
\end{equation}
for the searcher's position $x_n$, where $\xi_{\alpha}(n)$ is a set of random
variables sampled from a symmetric L{\'e}vy stable distribution with the
characteristic function $\exp\left(-\left|k\right|^{\alpha}\right)$. The time
step is chosen as $\delta t=0.001$ for LFs in all
cases but those which required higher time resolution. In that case $\delta t=
0.0001$ was used (figure \ref{ShortTimeExpansion}). While we keep the generalised
diffusion coefficient $K_{\alpha}$ of dimension $\mathrm{cm}^{\alpha}/\mathrm{sec}$
in our analytical results, we set it to unity in the numerical analyses. The noise
$\xi_{\alpha}(n)$ was computed following the method described in \cite{LevyGenerator}.
For LFs the time dependence is then simply obtained by adding the time step $\delta t$
to a counter at each jump. The set of landing points for LFs exactly corresponds to
the set of end-of-relocation points for LWs and, hence, the same simulation procedure
was used for both processes. For LWs, in turn, the parameter $\delta t$ is not related
to the time resolution any longer, but rather describes the width of the jump length
distribution. We thus use $K_{\alpha}\delta t=0.001$ for the first
passage of LWs and $K_{\alpha}\delta t=0.0005$ for the first
hitting problem. In order to compute the
time-dependent characteristics for
LWs the durations of relocations were calculated from a given relocation length via
the speed $v_0$ of the walk that we take as a constant value. In our simulation of
LWs the direction at the beginning of each relocation event is changed with
likelihood $\frac{1}{2}$ (that is, the searcher continues in the same direction
with probability $\frac{1}{2}$), as the jump lengths are taken from the symmetric
distribution of the $\xi_{\alpha}$ entries.

In all cases considered in this paper a searcher eventually crosses the boundary or
finds the target with unit probability. Hence, the first-passage and first-hitting
properties can be characterised by the properly normalised PDFs. We will denote them
as $\wp_{\mathrm{PF}}(t)$ and $\wp_{\mathrm{PW}}(t)$ for the first-passage of LFs
and LWs, respectively, and $\wp_{\mathrm{HF}}(t)$ and $\wp_{\mathrm{HW}}(t)$ for
the first-hitting of LFs and LWs.

\section{First-passage properties of L{\'e}vy flights and L{\'e}vy walks}

In this section we focus on the first-passage dynamics of LFs and LWs. First we analyse
the case of LFs. In the following subsection for LWs we compare the results
with the LF case.

\subsection{First-passage for L\'evy flights}

\begin{figure}
\centering
\includegraphics[width=11.2cm]{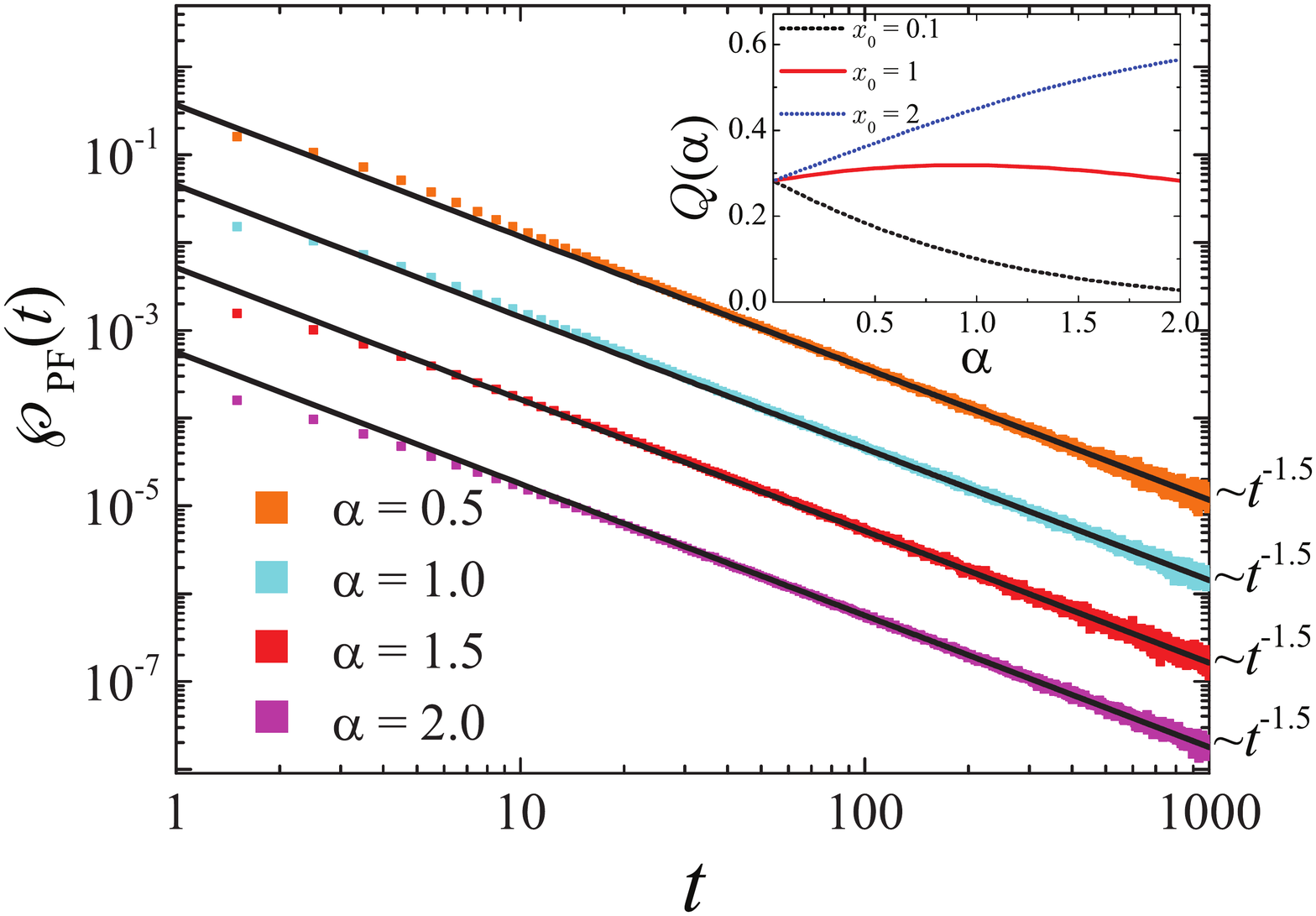}
\caption{PDFs of first-passage times for LFs illustrating the universal
Sparre Anderson-scaling (\ref{SparreAndersen}) valid in the long time limit.
Results for four different stable indices $\alpha$ are shown, as indicated
in the graph. Simulations data (coloured squares) were obtained from $N=10^7$
runs with initial position of the searchers $x_0=2$ and the generalised
diffusion coefficient $K_\alpha=1$. Black lines are obtained from
equation (\ref{SparreAndersen}). Note that for the better visual comparison
of the PDFs to the exact asymptotic scaling, their values are divided by a
factor of 10 for $\alpha=1$, of 100 for $\alpha=1.5$, and of 1000 for $\alpha
=2$. The inset shows the behaviour of the prefactor of the long-time dependence
$\simeq t^{-3/2}$, $Q(\alpha)=x_0^{\alpha/2}/[\alpha\sqrt{\pi}K_{\alpha}\Gamma
(\alpha/2)]$ from equation (\ref{SparreAndersen}), for different $x_0$ and with
$K_\alpha=1$. Larger $\alpha$ values lead to more pronounced differences in the
values of $Q(\alpha)$.}
\label{FPLFsLong}
\end{figure}

\begin{figure}\center
\includegraphics[width=11.2cm]{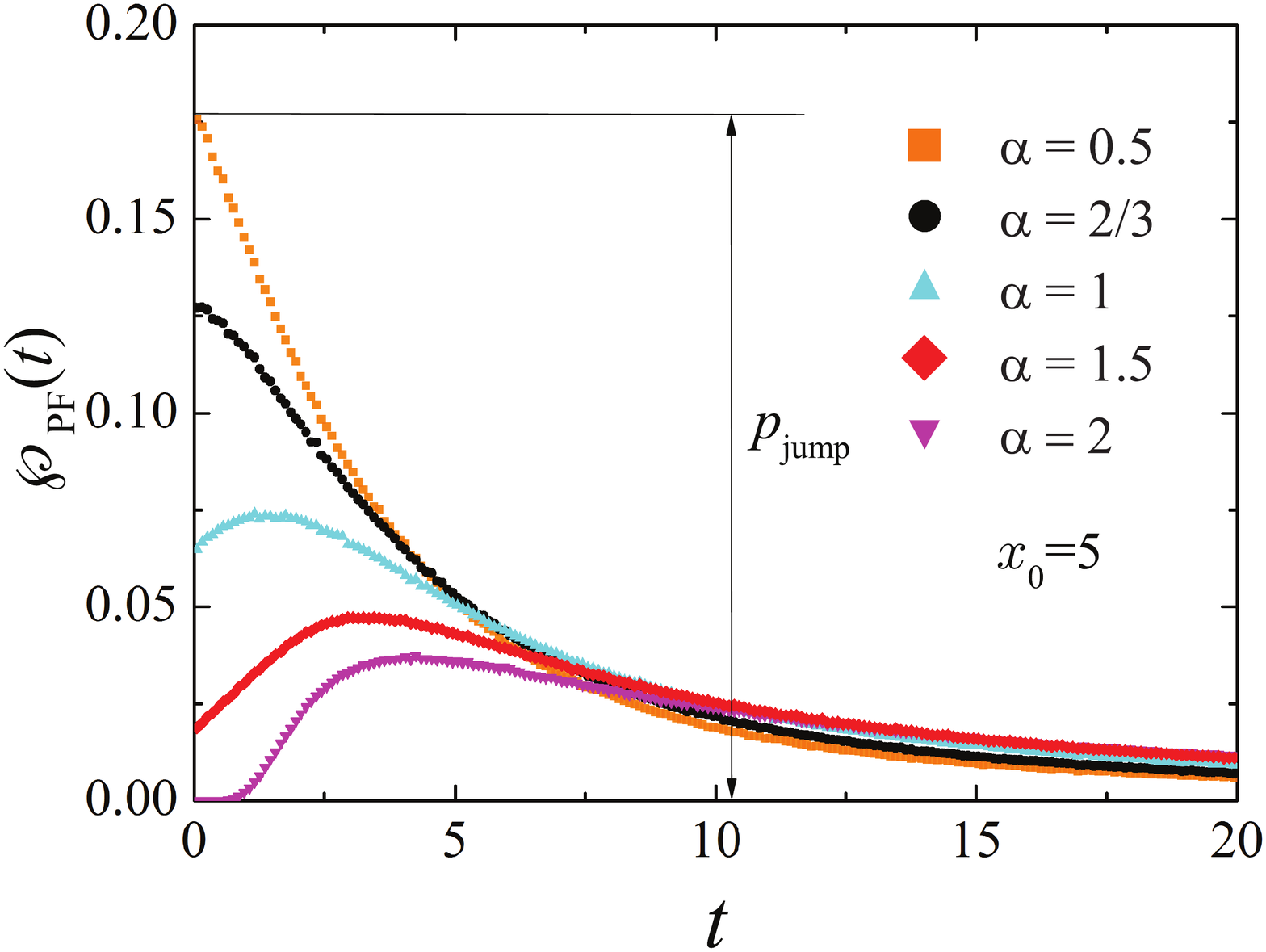}
\caption{Short time behaviour of the first-passage time PDF for LFs, demonstrating
that the change of the slope at $t=0$ occurs for $\alpha=2/3$ (for which we observe
a horizontal tangent). Here $p_\mathrm{jump}$ is the probability that the boundary
is immediately crossed with the first jump. Number of runs: $N=10^7$, initial
position: $x_0=5$.}
\label{FPLFsShortLin}
\end{figure}

The first-passage of LFs, due to their Markovian character and the symmetric jump
length distribution is necessarily characterised by the Sparre Andersen-scaling
\cite{Redner,JPA2003} in the long time limit. The analytical expression for this
limiting behaviour reads \cite{leapover2007}\footnote{Note that
we here consider the long time limit at fixed initial position. For a discussion
of the limiting behavior in a more general setting, where $x_0$ may diverge, we
refer the reader to \cite{satya}.}
\begin{equation}
\wp_{\mathrm{PF}}(t)\sim\frac{x_0^{\alpha/2}}{\alpha\sqrt{\pi K_\alpha}\Gamma
\left(\alpha/2\right)}t^{-3/2},
\label{SparreAndersen}
\end{equation}
in terms of the initial position $x_0$ and the stable index $\alpha$ of the jump
length PDF. Here and in the following, the symbol $\sim$ denotes asymptotic
equality, $\simeq$ means asymptotically equal up to a prefactor (scaling equality),
and $\approx$ means approximately equal. Figure \ref{FPLFsLong} illustrates that
the simulations results (coloured squares) are in perfect agreement with the
Sparre Andersen-scaling (\ref{SparreAndersen}) shown by the black lines for $\alpha$
values that are smaller and larger than unity. Note the shift by constant factors
between the different results.

Considering the corresponding short time behaviour in figure \ref{FPLFsShortLin}
one immediately realises that only for the case of Brownian motion ($\alpha=2$)
the PDF increases smoothly with time from the value zero at $t=0$. For LFs with
$\alpha<2$ the first-hitting PDF exhibits a non-zero value at $t=0$. For small
$\alpha$ values (see the curve for $\alpha=0.5$ in figure \ref{FPLFsShortLin})
the PDF decreases monotonically with time, while for larger $\alpha$ values an
initial increase is observed, leading to a maximum beyond which the PDF crosses
over to the long time Sparre Andersen-scaling. LFs with smaller $\alpha$ values have
a higher propensity for long jumps, while for $\alpha\to2$ the behaviour converges
to the known L{\'e}vy-Smirnov law for Brownian motion. The abrupt increase of
$\wp(t)$ at $t=0$ thus stems from LFs that directly overshoot the origin with their
first jump away from their initial position $x_0$. The associated probability
$p_\mathrm{jump}$ can be estimated from the survival probability of the searcher
(for the full derivation see \ref{pjumpAppendix}),
\begin{eqnarray}
p_\mathrm{jump}\approx\frac{K_\alpha\sin\left(\pi\alpha/2\right)\Gamma(\alpha)}{
\pi x_0^\alpha}.
\end{eqnarray}
For the $\alpha$ values considered in figure \ref{FPLFsShortLin} we obtain the
concrete values
\begin{equation}
p_\mathrm{jump}\approx\left\{\begin{array}{ll}\displaystyle\frac{K_{1/2}}{\sqrt{
2\pi x_0}}\approx0.178,&\alpha=\frac{1}{2}\\[0.4cm]
\displaystyle\frac{K_1}{\pi x_0}\approx0.064,&\alpha=1\\[0.4cm]
\displaystyle\frac{K_{3/2}}{2\sqrt{2\pi x_0^3}}\approx0.0178,&\alpha=\frac{3}{2}
\end{array}\right.,
\end{equation}
where we used $x_0=5$ and $K_\alpha=1$. These values are in perfect agreement
with the simulations data in figure \ref{FPLFsShortLin}.

Analytically one can also obtain the derivative of the PDF $\wp_\mathrm{PF}(t)$
for LFs in the short time limit (\ref{pjumpAppendix}). It turns out that the
monotonic decrease for small $\alpha$ values changes to the initial increase of
the PDF at the value $\alpha=2/3$, see figure \ref{FPLFsShortLin}.
We note that in \cite{sp_add} the "limited space displacement"
of the trajectory corresponding to the Cauchy-Lorentz distribution for relatively
low noise intensities at short times was discovered and analyzed.

\subsection{First-passage for L\'evy walks}
\label{fplw}

\begin{figure}
\centering
\includegraphics[width=11.2cm]{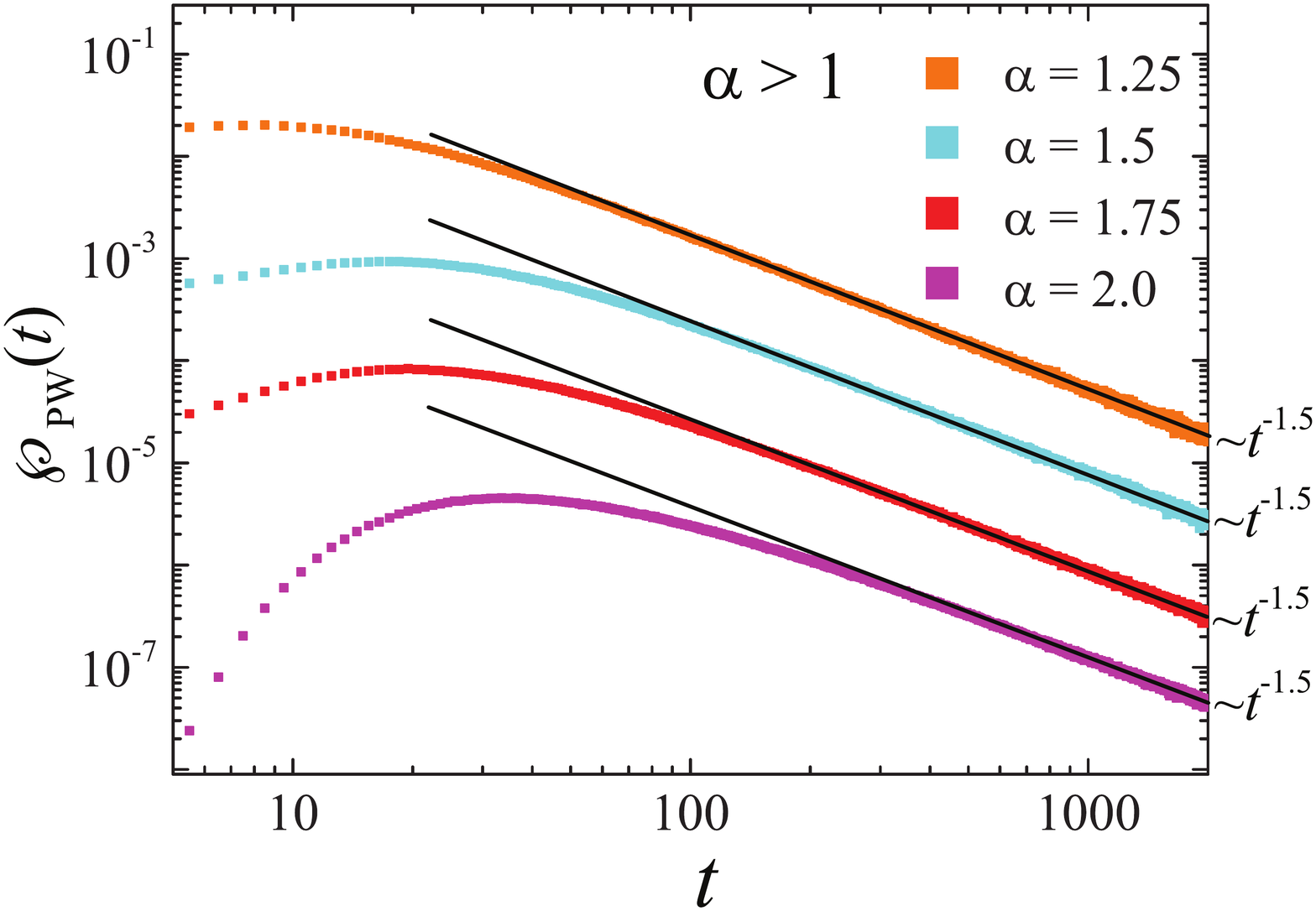}
\caption{Long time scaling behaviour of the first-passage PDF $\wp_{\mathrm{PW}}(t)$
of LWs for different $\alpha>1$. All curves nicely follow the predicted Sparre
Andersen-scaling (\ref{SparreAndersenLWs}). Number of runs: $N=10^7$, with initial
position $x_0=2$. Speed of the LW: $v_0=1$. Note that for the better visual
comparison of the PDFs their values are divided by factors of 10 for $\alpha=1.5$,
of 100 for $\alpha=1.75$ and of 1000 for $\alpha=2$.}
\label{FPLWsLong1}
\end{figure}

As laid out before, LWs spatiotemporally couple each relocation distance $x$ with a
time cost $t=x/v_0$. Nevertheless, starting from a L{\'e}vy stable jump length
distribution, the points of visitation of an LW are identical with those of an LF
with the same entries for the sequence of jumps---only the time counter for reaching
the points of visitation differs for both processes. Hence, many properties of LWs
can be understood from a subordination approach \cite{LeeWhitmore93}: In some sense,
to be specified below, an LW process can be considered as an LF with transformed
durations of individual jumps. The PDF for the relocation times for LWs follows from
the jump length PDF of LFs, as we show in \ref{JumpDistributionsLW}. From the point
of view of subordination one has to discriminate the two cases $\alpha>1$ and $\alpha
\le1$. Namely, for $\alpha>1$ the average duration of a jump is finite, and thus in
the limit of a large number of jumps the time characteristics of LWs and LFs will
only differ by a prefactor but should have the same scaling in time. From this
scaling argument we expect the Sparre Andersen-scaling
\begin{equation}
\label{SparreAndersenLWs}
\wp_{\mathrm{PW}}(t)\simeq t^{-3/2},\,\,\,\alpha>1.
\end{equation} 
to hold for the PDF $\wp_{\mathrm{PW}}(t)$ of the first-passage of LWs, as long as
$1<\alpha\le2$. Indeed, the simulations results presented in figure \ref{FPLWsLong1}
for 4 different values of $\alpha$ nicely corroborate the Sparre Andersen-$\frac{3}{
2}$ scaling (\ref{SparreAndersenLWs}).

The problem of escape of LFs and LWs for $\alpha>1$ was studied in
\cite{BartokEli2017}. The latter
paper also proposes a formula for the effective diffusion coefficient for LWs
expressed through the LW speed $v_0$, the scaling factor $\sigma_0$ or the
L{\'e}vy stable jump length PDF, and the stable index $\alpha$ (equation (55) in
\cite{BartokEli2017}). We derive this formula exactly in \ref{JumpDistributionsLW}
and \ref{LWdiffcoefDerivation}. Having an expression for the effective diffusion
coefficient $K^{\mathrm{LW}}_\alpha$ of LWs it seems pertinent to apply expression
(\ref{SparreAndersen}) to the LW dynamics. However, the numerical values obtained
in this vein do not coincide with the simulation results. The values from
the analytical results underestimate the simulations data in the limit of long $t$.
This discrepancy is caused by the differences in the \emph{short-time behaviour\/}
between LFs and LWs. Namely, until the front of an LW reaches the boundary, the
PDF of first-passage has a strictly zero value, due to the finite propagation front
of LWs \cite{RevModPhys2015}. Modelling LWs by a rescaled LF process obviously
neglects this important short-time feature. Hence a model LF process will exceed
the real LW PDF at short times and, for reasons of normalisation, underestimate the
PDF $\wp_{\mathrm{PW}}(t)$ at long times.

\begin{figure}
\centering
\includegraphics[width=11.2cm]{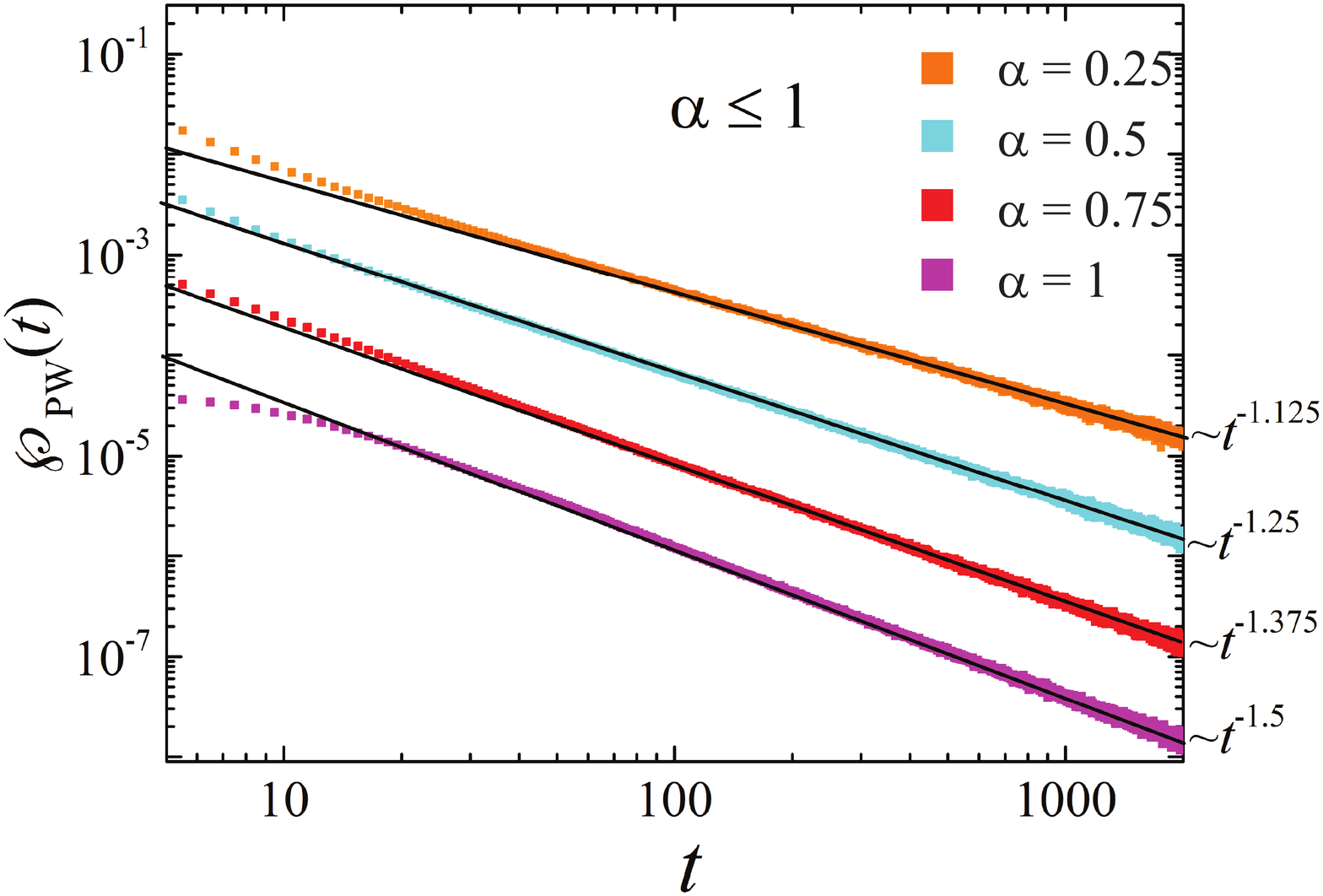}
\caption{Long time scaling of the first-passage PDF $\wp_{\mathrm{PW}}(t)$ of
LWs for different $\alpha\le1$. The fitted scaling exponents are in nice agreement
with the prediction of equation (\ref{fpt_lw}). The results are obtained from $N=10
^7$ runs with initial
position $x_0=2$, and relocation speed $v_0=1$. Note that for the better
visual comparison of the PDFs their values are divided by a factor of 10 for
$\alpha=0.5$, of 100 for $\alpha=0.75$ and of 1000 for $\alpha=1$.}
\label{FPLWsLong2}
\end{figure}

For LWs with $\alpha\le1$ the observed long-time scaling deviates from the
Sparre Andersen-scaling, as evidenced by the simulations data shown in figure
\ref{FPLWsLong2}. The fitted scaling exponents are consistent with the long
time scaling
\begin{equation}
\label{fpt_lw}
\wp_{\mathrm{PW}}(t)\simeq t^{-\alpha/2-1},\,\,\,\alpha\le1
\end{equation}
considered in \cite{Korabel2011} which can also be rationalised in subordination
terms. Indeed, the survival probability of an LF in the case of a first-passage
scenario is inversely proportional to the square root of the number $n$ of flights,
that is, $\mathscr{S}_{\mathrm{LF}}(n)\simeq n^{-1/2}$. For $\alpha<1$ the number
of jumps scales like $n\simeq t^{\alpha}$, such that $\mathscr{S}_{\mathrm{LF}}(t)
\sim t^{-\alpha/2}$. This scaling is equivalent to that obtained from the
subdiffusive fractional diffusion equation \cite{RalfPhysA2000}. Since the
first-passage time PDF is the negative of the first derivative of the survival
probability, we get $\wp_{\mathrm{PW}}(t)\simeq
t^{-\alpha/2-1}$. A strict derivation of this exponent for LWs can be found in
\cite{artuso}, where the general result for $\wp_{\mathrm{PW}}(s)$ is shown to be a
function of the step length distribution in Laplace space. This general expression
produces both the Sparre Andersen-scaling for $\alpha>1$ and the exponent $-1-\alpha/2$
for $\alpha<1$.

\begin{figure}
\centering
\includegraphics[width=11.2cm]{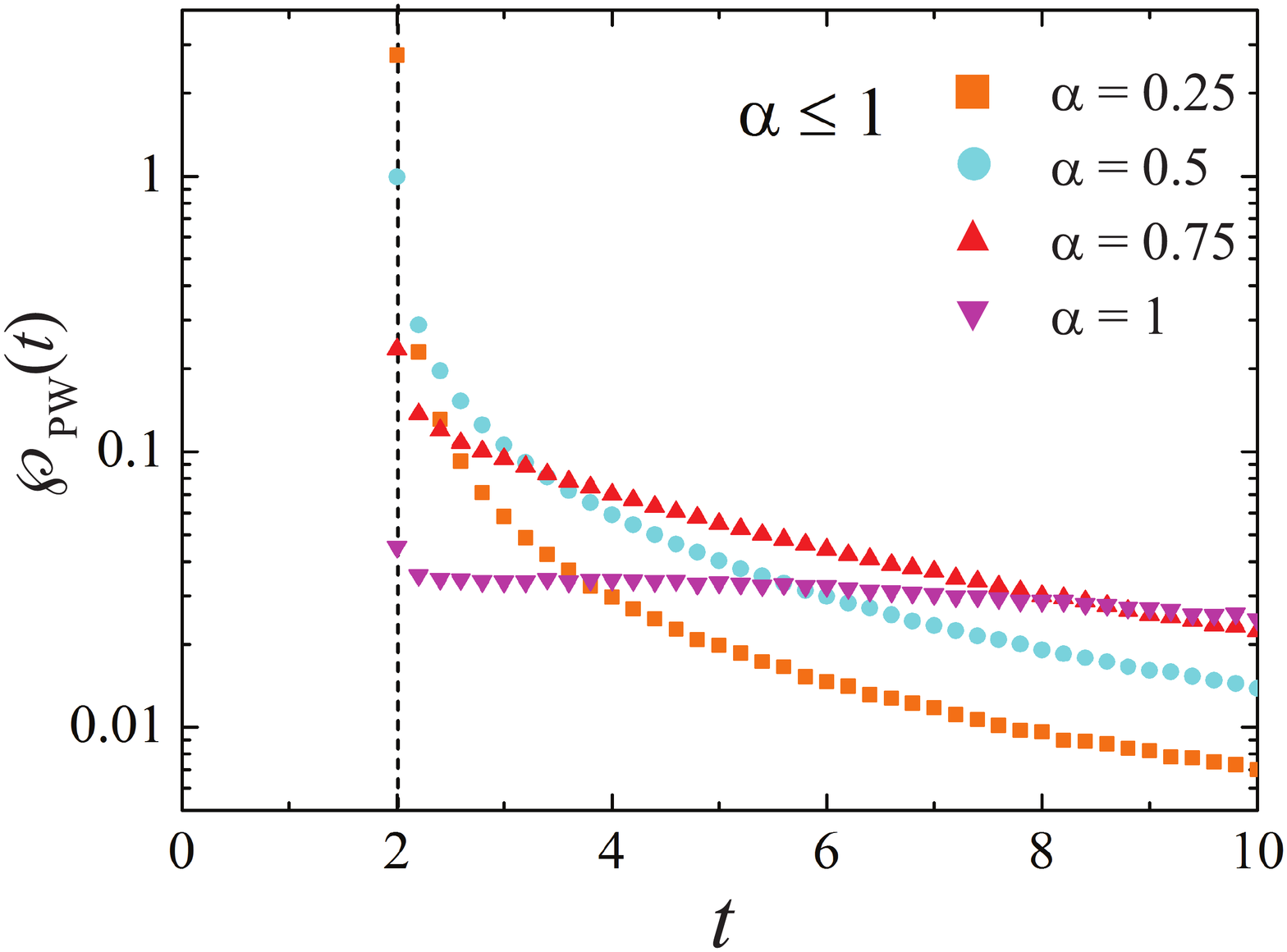}
\caption{Short-time behaviour of the first-passage PDF of LWs for different
$\alpha\le1$. The results are averaged over $N=10^7$ runs with initial position
was $x_0=2$ and LW speed $v_0=1$.}
\label{FPLWsShort1}
\end{figure}

\begin{figure}
\centering
\includegraphics[width=11.2cm]{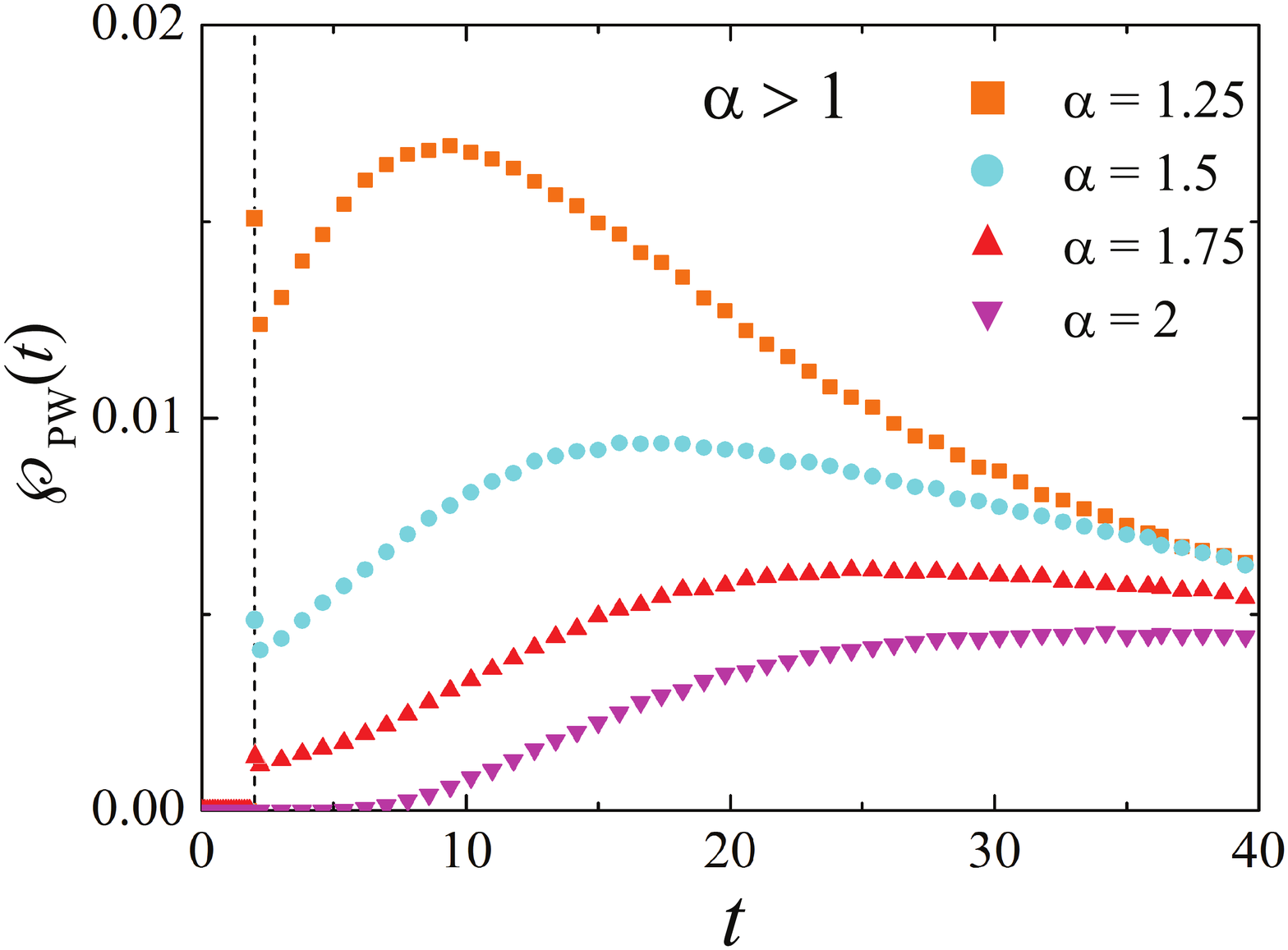}
\caption{Short-time behaviour of the first-passage PDF of LWs for different
$\alpha>1$. The results are averaged over $N=10^7$ runs, the initial position 
was $x_0=2$, and the LW speed is $v_0=1$. The bin size was chosen as $0.1$.}
\label{FPLWsShort2}
\end{figure}

The short-time limit of the first-passage time PDFs of LWs shown in figures
\ref{FPLWsShort1} and \ref{FPLWsShort2} is noticeably different from the
behaviour of LFs (compare figure \ref{FPLFsShortLin}). Namely, for LWs the
spatial spreading is limited by a front which travels with the constant speed
$v_0$, while for the LF case the PDF $\wp_{\mathrm{PF}}(t)$ has non-zero values
at any non-zero time. Before the front of an LW reaches the origin the PDF of
first-passage times is identically zero. At exactly $t=x_0/v_0$ we observe a
jump in the value of the first-passage PDF. This jump corresponds to all those
walkers which did not change the direction even once since the process started.
Then, similarly to LFs, the function decays monotonically for small $\alpha$ or
has an intermediate maximum.

We notice that in the limit $\alpha=2$ of Gaussian relocations the resulting
motion performed with a constant speed is similar to the model of immortal
creepers \cite{Abad2015}. For this type of motion the first-passage behaviour
was studied. However, in the model considered in \cite{Abad2015} the creepers
change direction with a waiting time PDF $\psi_{\mathrm{creeper}}(t)=\omega e^{
-\omega t}$, where $\omega$ is the characteristic turning frequency. In our case
one has to compute the distribution of relocation times from the Gaussian
characteristic function of the process. For $\alpha=2$ in our case (compare
\ref{JumpDistributionsLW}),
\begin{equation}
\psi(\tau)=\frac{1}{\sqrt\pi}\frac{v_0}{\sigma_0}\exp\left(-\frac{v_0^2\tau^2}{
4\sigma^2_0}\right),\,\,\,\left<\tau\right>=\frac{2\sigma_0}{v_0\sqrt\pi}.
\end{equation}
Hence the equation (56) for the survival probability derived in \cite{Abad2015}
is not directly applicable in our case.

Generally, the jump of $\wp_{\mathrm{PW}}(t)$ occurs at the very moment when the
propagation front, the fraction of particles having moved the distance $vt$ without
direction changes, passes through the boundary. This front corresponds to a delta
peak with decreasing amplitude, moving with the wave variable $|x|\pm v_0t$
(\ref{JumpPWderivation}). Formally, that is, the value of the PDF at the time when
this peak reaches the boundary is infinite, $\wp_{\mathrm{PW}}(t_0=x_0/v_0)\to
\infty$, as the survival probability changes as a step function.\footnote{The
prefactor of the $\delta$-function can be computed analytically as shown in
\ref{JumpPWderivation}}. In the simulations the PDF values are obtained from
collecting crossing events in finite bins whose size represents the time resolution.
Hence, the finite numerical values at $t_0$ are an artefact of the binning,
and a finer mesh will produce larger values of $\wp_{\mathrm{PW}}(t)$ (at $t=2$ for
the simulations parameters used for figure \ref{FPLWsShort2}). We verified that this
is indeed the case, as demonstrated by table \ref{tab1}.

The difference in behaviour of the first-passage PDF between the cases $\alpha\le1$
and $\alpha>1$ as shown in figures \ref{FPLWsShort1} and \ref{FPLWsShort2} can be
explained by the difference in the shape of the LW propagators (compare figure
\ref{supp_pdf}). In fact, for $\alpha>1$ the position PDF of LWs has
a front (smaller spikes in figure \ref{supp_pdf}) producing the jump in the
first-passage PDF, then the main, bell-shaped part of the PDF arrives, and the
first-passage PDF grows gradually. For $\alpha<1$, in contrast, the fronts
are the only maxima in the position PDF, and thus the PDF of first-passage has a
non-zero value and then decays monotonically.

\begin{table}
\centering
\begin{tabular}{|c|c|c|}
\hline
Bin size & $\alpha=0.5$ & $\alpha=1.5$ \\\hline
$10^{-1}$ & 0.998 & 0.0049 \\\hline
$10^{-2}$ & 3.232 & 0.0011 \\\hline
$10^{-3}$ & 10.52 & 0.0075 \\\hline
$10^{-4}$ & 35.16 & 0.718 \\\hline
\end{tabular}
\caption{Values of the LW first passage $\wp_{\mathrm{PW}}(t)$ at $t=2$ for
two different $\alpha$ values and different bin sizes.}
\label{tab1}
\end{table}

\section{First-hitting properties of L{\'e}vy flights and L{\'e}vy walks}

In order to observe first-hitting events in a one-dimensional setting, we necessarily
need to require that the first absolute moment $\langle|x|\rangle$ of the jump length
PDF exists, corresponding to the requirement $\alpha>1$ \cite{JPA16,JStatMech14}. For
the opposite case of $0<\alpha<1$ the associated first-hitting time PDF vanished
identically to zero. We do not consider the latter case here. For LWs an event of
first-hitting in our setting occurs when the end point of a relocation with speed
$v_0$ hits the target.

\subsection{First-hitting properties of L\'evy flights with $\alpha>1$}

The probability of first-hitting (see figure \ref{PassageVSArrival}) clearly depends
on the exact target size \cite{AlekseiBartekJPA16}. Here we will concentrate on the
case of point-like targets. Detailed studies of the first-hitting properties of LFs
were presented in \cite{PNAS14,JStatMech14,JPA2003}. Analytical derivations are
based on the fractional Fokker-Planck equation \cite{report,epl} with a sink term,
\begin{equation}
\label{SinkFFPE}
\frac{\partial f(x,t)}{\partial t}=K_{\alpha}\frac{\partial^{\alpha}f(x,t)}{\partial
\left\vert x \right\vert^{\alpha}}-\wp_{\mathrm{HF}}(t)\delta(x),
\end{equation}
where the fractional derivative is defined in terms of its Fourier transform,
$\int_{-\infty}^{\infty}\exp(ikx)[\partial^{\alpha}f(x,t)/\partial|x|^{\alpha}]dx=
-|k|^{\alpha}f(k,t)$, where $f(k,t)$ is the Fourier transform of $f(x,t)$.
Equation (\ref{SinkFFPE}) can be easily modified to include an external drift
\cite{JStatMech14}, to the case of multiple point-like targets \cite{EPJB17}, or
to include an additional Brownian or L{\'e}vy component with different stable index
$\alpha'$ \cite{JPA16,EPJB17}. We consider here a perfectly absorbing sink. The
case of a finite absorption strength can be found in \cite{Janakiraman17}.

Equation (\ref{SinkFFPE}) can be solved in Laplace space for the initial condition
$f(x,0)=\delta(x-x_0)$ \cite{JPA2003}, yielding
\begin{equation}
\wp_{\mathrm{HF}}(s)=\frac{\displaystyle{\int_{-\infty}^{\infty}dk\frac{e^{ikx_0}}{
s+K_\alpha\left\vert k \right\vert^{\alpha}}}}{\displaystyle{\int_{-\infty}^{\infty}
dk\frac{1}{s+K_\alpha\left\vert k \right\vert^{\alpha}}}},
\label{twointsolution}
\end{equation}
From inverse Laplace transform of this expression we obtain a new result for the
first-hitting time PDF in time (see \ref{LongTimeAF}),
\begin{eqnarray}
\wp_{\mathrm{HF}}(t)=\frac{\alpha\sin\left(\pi/\alpha\right)K_\alpha^{1/\alpha}}{
\pi}t^{1/\alpha-1}\int_0^\infty dk\cos(kx_0)E_{1,1/\alpha}(-K_\alpha k^{\alpha}t),
\label{LFFApdftime}
\end{eqnarray}
where $E_{1,1/\alpha}(-K_\alpha k^{\alpha}t)$ is a two-parameter Mittag-Leffler
function, which can be defined by its series expansions \cite{erdelyi}
\begin{equation}
E_{\alpha,\beta}(-z)=\sum_{k=0}^{\infty}\frac{(-z)^k}{\Gamma(\beta+\alpha k)}\sim
\sum_{k=1}^{\infty}\frac{(-1)^{k+1}}{z^k\Gamma(\beta-\alpha k)},\,\,\, z>0
\end{equation}
around zero and infinity, respectively. From the latter expansion we immediately
recover the well-established power law asymptotic for the first-hitting time PDF
of LFs \cite{JPA2003},
\begin{equation}
\label{lt}
\wp_{\mathrm{HF}}(t)\simeq t^{1/\alpha-2}
\end{equation}
at long times (compare equation (\ref{longAFanalyt}) and its derivation in
\ref{LongTimeAF}). We here also derive the short time scaling law
\begin{equation}
\label{st}
\wp_{\mathrm{HF}}(t)\simeq t^{1/\alpha}
\end{equation}
in \ref{ShortTimeAF}. The full expression (\ref{LFFApdftime}) can be evaluated
numerically to plot the analytical solution of (\ref{SinkFFPE}).

To determine the first-hitting time PDF from simulations of a searcher jumping
according to a L{\'e}vy stable jump length distribution, we need to endow the
target with a finite size $d$. This size needs to be large enough to guarantee
sufficiently many events for proper statistics, however, it should be small
enough to vouchsafe the point-like character of the target and thus warrant
consistency with equation (\ref{SinkFFPE}). This correct size also depends on
the time resolution and the time cutoff of the simulation run but not on the
number of runs (see Appendix G). As a criterion for the choice of the target size
we checked agreement with both the expected long-time power law (\ref{lt}) and the 
short-time power law (\ref{st}). We found that for the integration time step
$\delta t=0.001$ and the time limit $10^3$ for every run the proper target sizes
are $d=0.06$ for $\alpha=2$ (Brownian
motion), $d=0.035$ for $\alpha=1.75$, $d=0.014$ for $\alpha=1.5$, and $d=0.003$
for $\alpha=1.25$. In particular we note here that this question of the finite
target size is not limited to L{\'e}vy stable motion, but is also an issue for
regular Brownian motion.

\begin{figure}
\centering
\includegraphics[width=11.2cm]{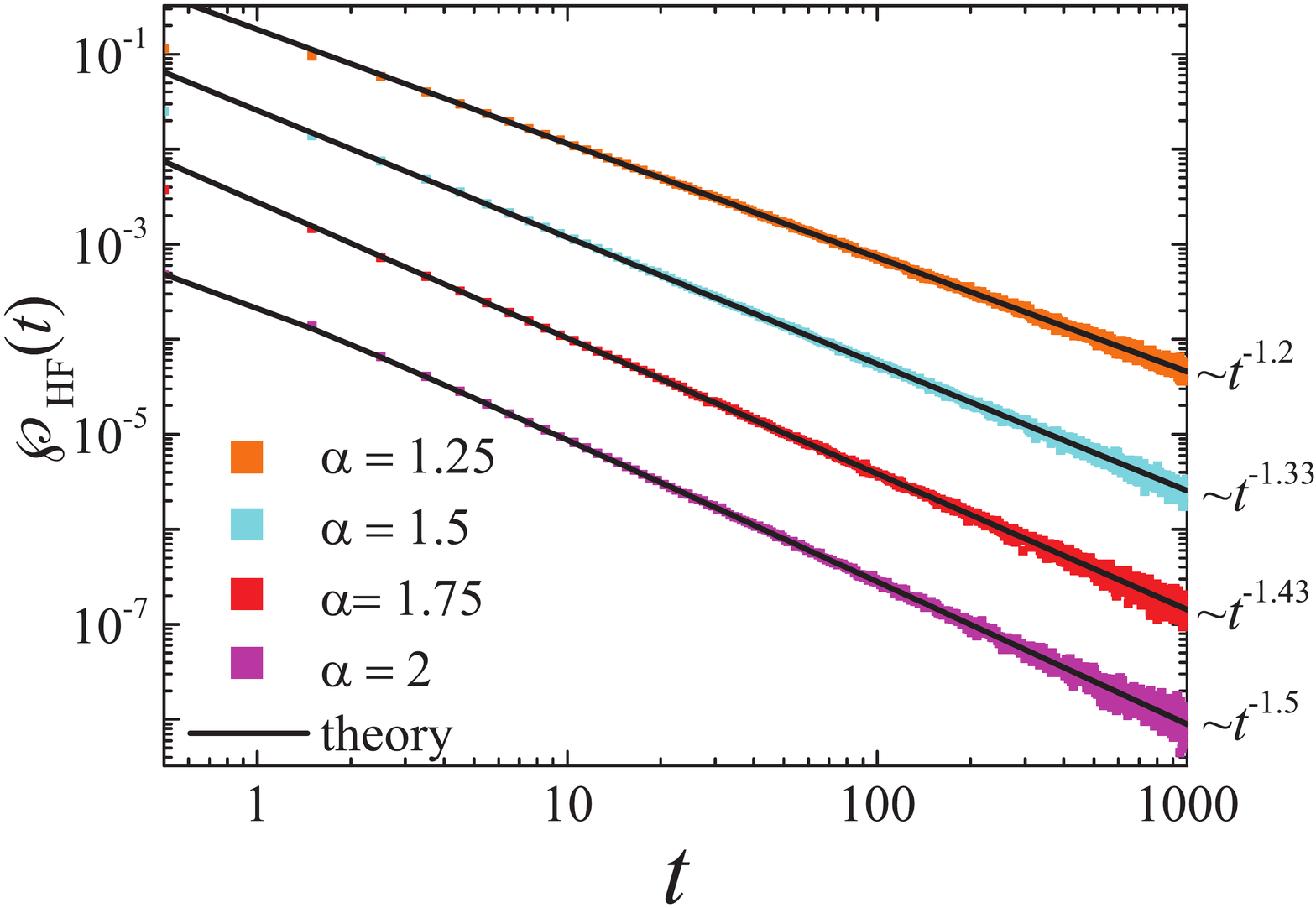}
\caption{First-hitting time PDF for LFs with different $\alpha$ illustrating the
long time scaling law (\ref{lt}), here represented by the black lines with slopes
predicted by the analytical expression. The coloured symbols depict the simulations
results. Note that for better visual comparison the PDFs are divided by a factor
of 10 for $\alpha=1.5$, of 100 for $\alpha=1.75$, and of 1000 for $\alpha=2$. We
chose $K_
\alpha=1$.}
\label{FALFsLong}
\end{figure}

\begin{figure}
\centering
\includegraphics[width=11.2cm]{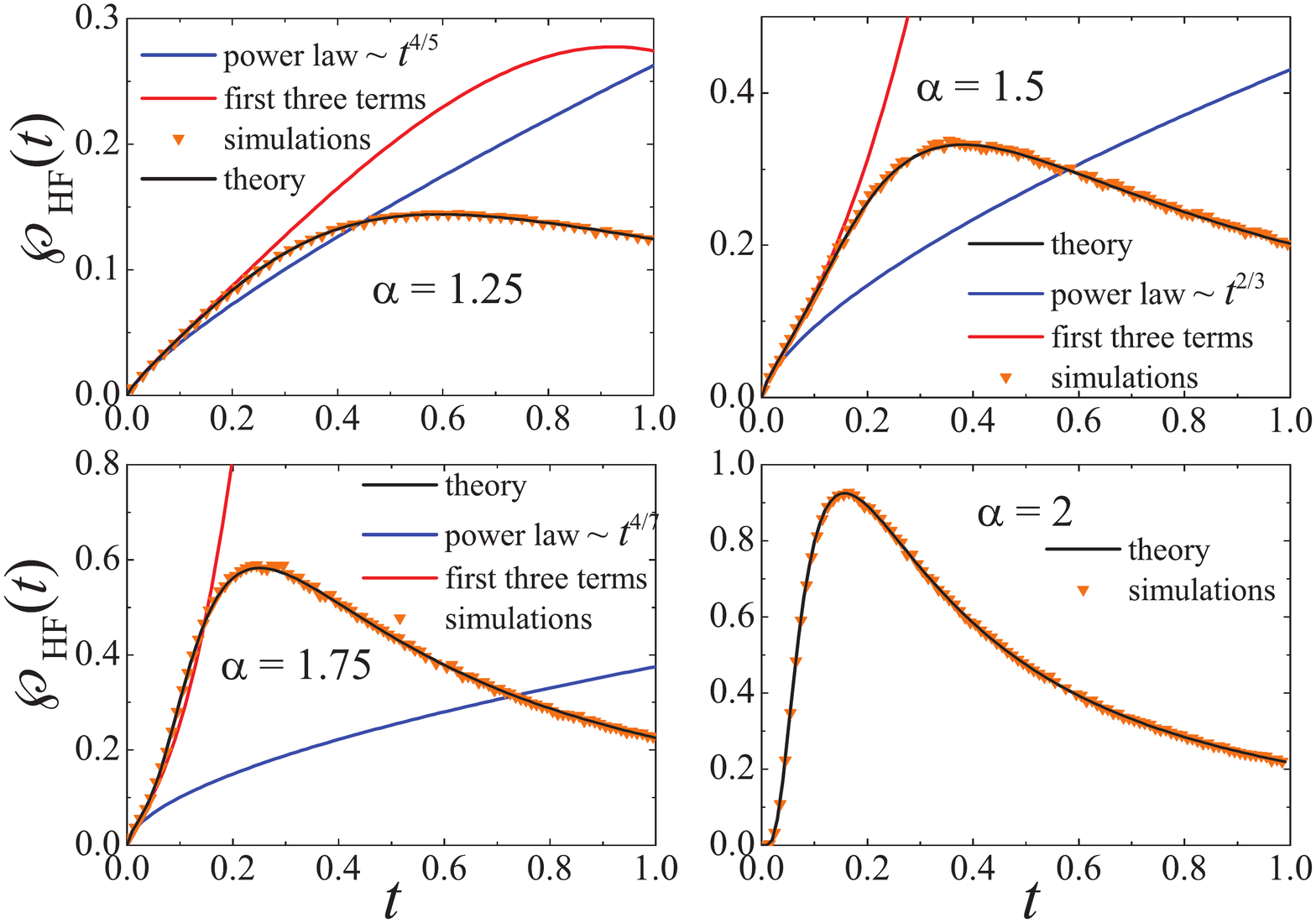}
\caption{Short-time asymptotic for the first-hitting time PDF of LFs. The black
lines are computed from the full expression (\ref{LFFApdftime}), the blue lines
show the power-law (\ref{st}) with the exact prefactor as given by the first term
in expansion (\ref{expansion1}), equation (\ref{I1}). The red lines shows the sum
of the first three terms of the expansion, combining equations (\ref{I1}), (\ref{I2}).
and (\ref{I3}). Parameters: $K_\alpha=1$ and $x_0=1$.}
\label{ShortTimeExpansion}
\end{figure}

Figure \ref{FALFsLong} shows the long-time behaviour of the first-hitting time PDF
for LFs. The PDFs obtained from simulations (coloured symbols) are well fitted by
the theoretical result (\ref{longAFanalyt}). In figure \ref{ShortTimeExpansion} we
illustrate the corresponding short-time behaviour for the first-hitting time PDF.
The black curves show the exact analytical solution computed from equation
(\ref{LFFApdftime}). Orange triangles are the simulation results. The correspondence
between the analytical expression and the simulations is excellent. The blue curves
correspond to the first term in the expansion (\ref{expansion1}) proportional to
$t^{1/\alpha}$, which apparently works better for smaller $\alpha$. For $\alpha=2$
all coefficients in expansion (\ref{expansion1}) become equal to zero, that is, it
does not work for the Brownian case. Generally, the terms of the expansion depend
on time as $t^{k+1/\alpha}$ with $k$ being the degree of the expansion. Once more
terms of the expansion are added the quality of the approximation improves
substantially (red curves in figure \ref{ShortTimeExpansion}).

\subsection{First-hitting probability for L\'evy walks with $\alpha>1$}

\begin{figure}
\centering
\vspace*{0.2cm}
\includegraphics[width=11.2cm]{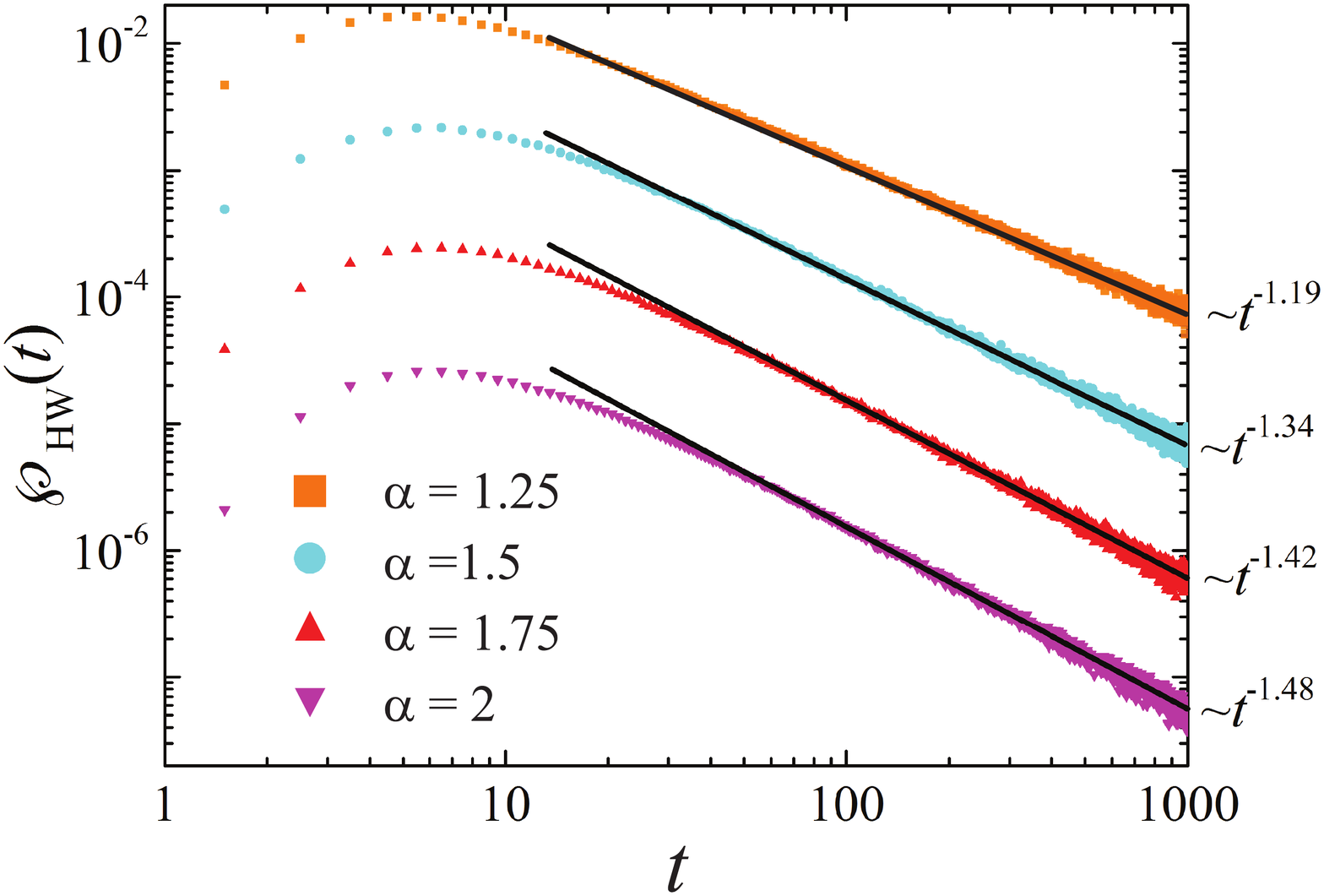}
\caption{Long-time behaviour of the first-hitting time PDF of LWs. The exponents
derived from the simulations coincide with those for LFs calculated above,
within the error margin. The target sizes were the same as for the LFs.
Parameters: $x_0=1$ and $v_0=1$.}
\label{FALWsLong} \end{figure}

\begin{figure}
\centering
\includegraphics[width=7.6cm]{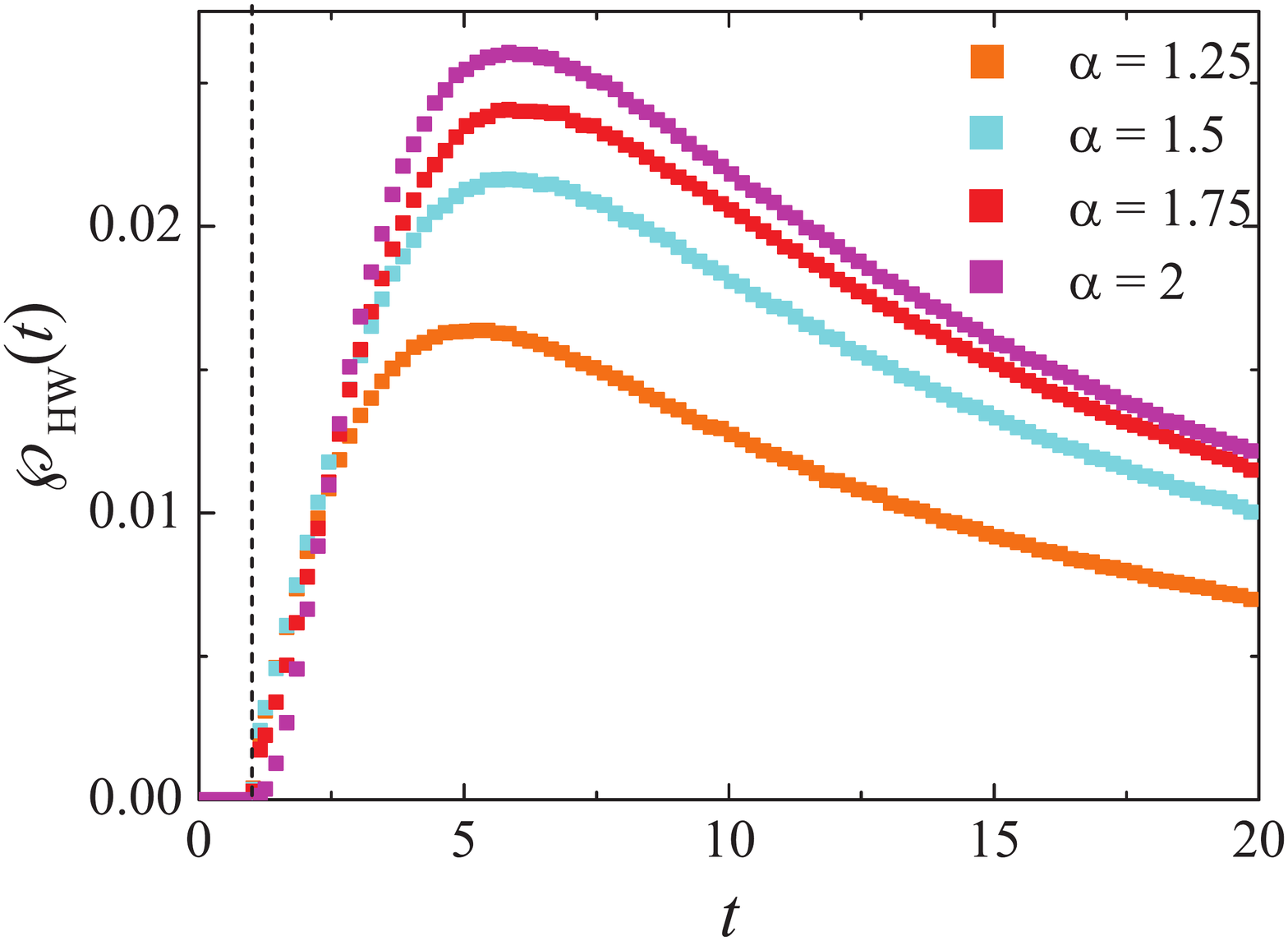}
\includegraphics[width=7.6cm]{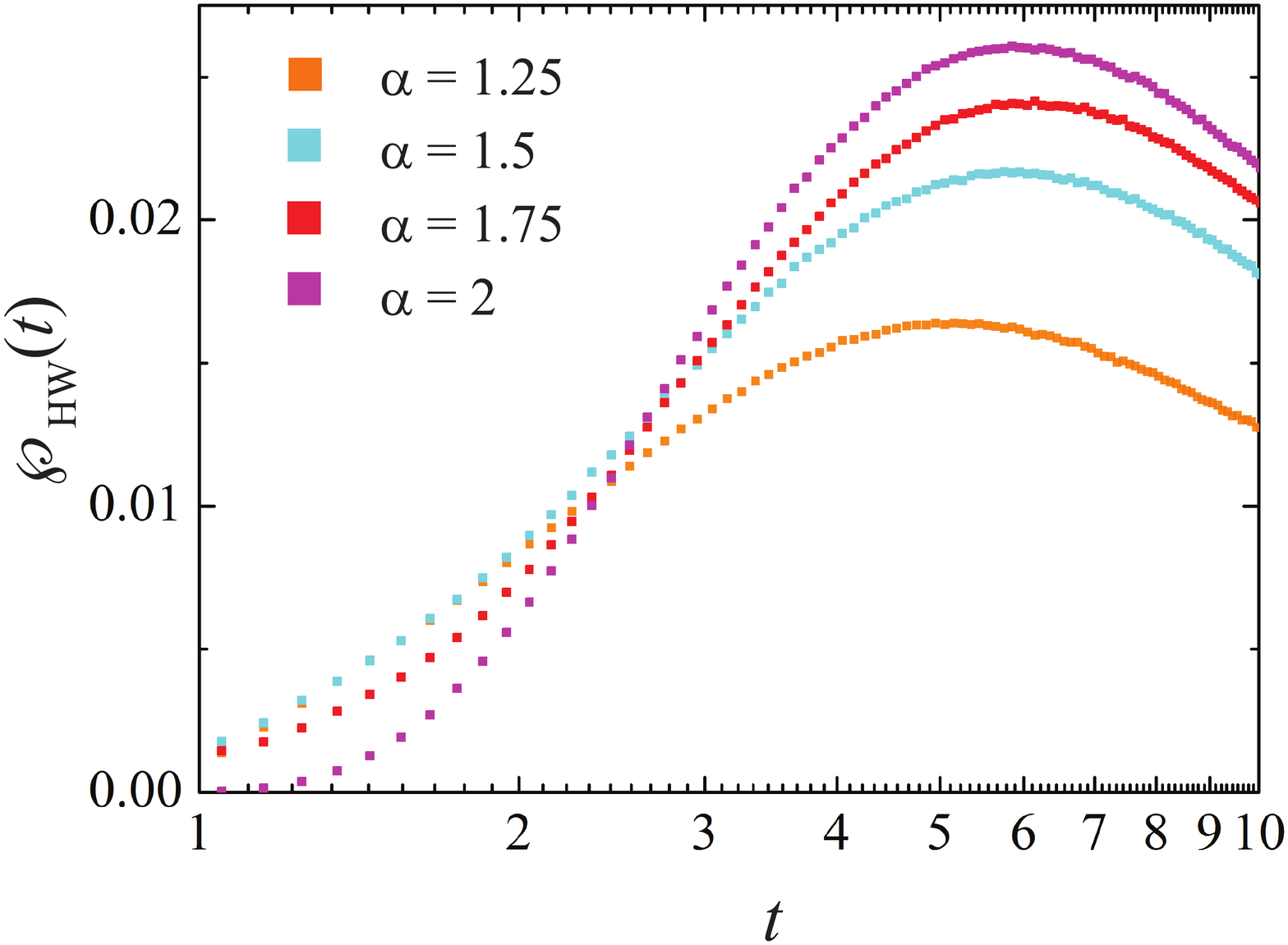}
\caption{Short-time behaviour of first-hitting time PDF of LWs. Parameters:
$x_0=1$ and $v_0=1$. Left: linear scale. Right: logarithmic
scale.}
\label{FALWsShort}
\end{figure}

In figure \ref{FALWsLong} we show
the numerical data for this case in the limit of long times. The numerically
determined scaling exponents nicely coincide with those obtained analytically
and numerically for LFs (compare figure \ref{FALFsLong}). The reason for this
similarity is the same as for the case of first-passage: In the long time limit
for $\alpha>1$ both processes have the same scaling behaviour due to the existence
of a finite scale of the jumps in the processes. The only difference is that the
prefactors in both cases may differ, corresponding to a renormalisation of the
mean step time.

Naturally, the short-time properties of the first-hitting behaviour depicted in
figure \ref{FALWsShort} for LWs differ from the corresponding shape for LFs, in
analogy to our observations for the first-passage case. Namely, no first-hitting
event can occur before the LW front reaches the target. Then a jump in the value
occurs at $t=x_0/v_0$. Similar to our discussion of the first-passage time PDF
we attempted to insert an effective diffusion coefficient $K_\alpha^{\mathrm{LW}}$
in the known long-time expressions for LFs, equation (\ref{longAFanalyt}). Given
that this did not succeed for the first-passage scenario, it is not surprising
that an effective description in terms of LFs cannot approximate the LW behaviour.

\section{Conclusions}

LFs and LWs are broadly used models for efficient random search processes.
In this paper we systematically analysed and compared the first-passage and
first-hitting properties for two different models of L{\'e}vy motion, namely,
LFs and LWs. We demonstrated that for $\alpha>1$ the results of these two models
are qualitatively identical at long times due to the finite average length
of a relocation. The situation drastically changes for $\alpha<1$ when
the scaling of the PDFs heavily depends on the exact model (see the summary
in table \ref{tab}). This difference does not come as a surprise in view of the
strong difference of the propagators of the two models, particularly, the finite
propagation front of LWs. Nevertheless, having quantitative data for the associated
first-passage and first-hitting time PDFs is valuable for any concrete analysis of
search models based on LFs and LWs.

\begin{table}
\begin{center}
\label{table1}
\renewcommand{\arraystretch}{2}
\begin{tabular}{|c|c|c|}
\hline
\textbf{} & \textbf{First-passage} & \textbf{First-hitting}\\
\hline
\multicolumn{1}{|c|}{\textbf{Brownian motion}} & \multicolumn{2}{l|}{$\wp_{
\mathrm{PBM}}(t)=\wp_{\mathrm{HBM}}(t)=\frac{x_0}{\sqrt{4\pi K_2t^3}}\exp
\left(-\frac{x_0^2}{4K_2 t}\right)$}\\
\hline     
\textbf{Brownian creepers} & $\wp_{\mathrm{PBC}}(t)\simeq t^{-3/2}$ &
$\wp_{\mathrm{HBC}}(t)\simeq t^{-3/2}$\\
\hline     
\textbf{L{\'e}vy flights} (1\textless$\alpha<2$) & $\wp_{\mathrm{PF}}(t)\simeq
t^{-3/2}$ & $\wp_{\mathrm{HF}}(t)\simeq t^{1/\alpha-2}$\\
\hline
\textbf{L{\'e}vy flights} ($\alpha\leq1$) & $\wp_{\mathrm{PF}}(t)\simeq t^{-3/2}$
& $\wp_{\mathrm{HF}}(t)=0$\\
\hline
\textbf{L{\'e}vy walks} ($1<\alpha\le2$) & $\wp_{\mathrm{PW}}(t)\simeq t^{-3/2}$
& $\wp_{\mathrm{HW}}(t)\simeq t^{1/\alpha-2}$\\
\hline
\textbf{L{\'e}vy walks }($\alpha\le1$) & $\wp_{\mathrm{PW}}(t)\simeq t^{1-\alpha/2}$
& $\wp_{\mathrm{HW}}(t)=0$\\
\hline
\end{tabular}
\end{center}
\caption{Comparison of the long-time asymptotes of the first-passage and first-hitting
PDFs for LFs and LWs.}
\label{tab}
\end{table}

While our findings are consistent with the results for the escape from an interval
established in \cite{BartokEli2017}, we observed that the exact short-time behaviour
of LFs and LWs influences the whole first-passage and first-hitting time PDFs, and
particularly alters the amplitude factor of the long-time scaling behaviour of LWs
with an effective diffusion coefficient. In this short-time limit the PDF of the
first-passage time of LFs jumps instantly to a finite value, which is not typical
for the first-hitting of LFs. In the case of LWs the short time behaviour is defined
by the passage of the propagation front through the boundary or the target. Strictly
speaking the front passes instantly through any coordinate. Hence, the value of the
PDF should be formally infinite at that instant. Obviously, this feature is
smoothened in simulations and real observations due to finite binning, but it reveals
itself through a strong dependence of the value of the PDF at this instant on the
time resolution of the simulations. In essence we showed that it is important to
know the entire first-passage or first-hitting PDF in order to reach fully
quantitative conclusions.

It will be interesting to extend the results reported herein to higher
dimensions and/or multiple target scenarios. Higher-dimensional LFs and LWs can
be defined in different ways \cite{Marek2012,EliMult}, and it is not a priori clear
whether the difference in definitions modifies the behaviour for the
first-passage and first-hitting time PDFs. In the case of multiple targets, for
instance, disks distributed randomly in a plane, the associated first-passage
and first-hitting time PDFs to locate these disks would move this problem closer
to biological reality. One might think of, as an example, a bee searching for
flowers, a situation amenable to experiments \cite{LICCK12}. Measuring the
first-passage and first-hitting time PDFs may help to put the LF foraging
hypothesis on more rigorous theoretical grounds, which is a long-standing
open problem.

Another interesting direction to study is the combination of
L{\'e}vy-type processes with resetting, as studied, for instance, for mean first
passage and arrival times in \cite{schehr}.
Moreover, it will be interesting to explore in more detail
the impact of a bias on the first-passage and first-hitting time PDFs
similar to the analyses in \cite{PNAS14,JStatMech14}. Biologically this would
relate to the problem of chemotaxis when cells migrate according to a chemical
concentration gradient to find inflammation sites \cite{cell_chemo}.
Similarly, extensions of the present results in general
external potential fields \cite{sp1} should be studied, including phenomena
such as barrier crossing, stochastic resonant activation, and noise-enhanced
stability phenomena \cite{sp2}. Further applications to pursue include population
dynamics and biophysical models \cite{sp3}, as well as the modelling of the dynamics
in Josephson junctions \cite{sp4}, to mention but a few stray examples.
We moreover
note that it should be analysed how distributed transport parameters (such as
$K_{\alpha}$) may modify the results for first-passage and first-hitting,
similar to what happens for Brownian processes \cite{vittoria}. Finally, it
will be interesting to study more complex, many-body scenarios such as hunting
in flocks \cite{glebmaria}.

\appendix

\section{Short-time behaviour of the first-passage of L{\'e}vy flights}
\label{pjumpAppendix}

The finite value $p_{\mathrm{jump}}$ of the jump in the short-time behaviour of LFs
(figure \ref{FPLFsShortLin}) can be estimated by computing the survival probability
at short times by using the asymptotic expression for large $x$. For the purpose of
this derivation we assume that the starting position is at $x=0$ while the boundary
is at $x=x_0$, which is physically identical to our original setting. The survival
probability at short times reads
\begin{equation}
\mathscr{S}(t)=\int_{-\infty}^{x_0} f_\alpha(t,x)dx,
\end{equation}
where the symmetric L{\'e}vy stable PDF in the limit $|x|\to\infty$ is
\cite{ch1,leapover2007}
\begin{equation}
f_\alpha(t,x)\sim C_1(\alpha)\frac{K_\alpha t}{|x|^{1+\alpha}},\,\,\,
C_1(\alpha)=\frac{1}{\pi}\sin(\pi\alpha/2)\Gamma(1+\alpha).
\end{equation}
The first-passage time density is the negative derivative of the survival probability,
\begin{equation}
\wp_{\mathrm{PF}}(t)=-\frac{d\mathscr{S}(t)}{dt}.
\end{equation}
Conversely,
\begin{equation}
1-\mathscr{S}(t)=\int_{x_0}^\infty f_\alpha(t,x)dx=\frac{C_1(\alpha)K_\alpha t}{
\alpha x_0^\alpha}.
\end{equation}
Hence
\begin{equation}
p_{\mathrm{jump}}=\wp_{\mathrm{PF}}(t\rightarrow0)=\frac{K_\alpha\sin(\pi\alpha/2)
\Gamma(\alpha)}{\pi x_0^\alpha}.
\end{equation}

In order to estimate the slope at $t=0$ we will compute whether the values of
the PDF increase or decrease at short times. As we established above the
probability to cross a boundary within an infinitesimal time $\Delta t_1$ reads
\begin{equation}
p_1=1-\mathscr{S}(\Delta t_1)=\int_{x_0}^\infty f_\alpha(\Delta t_1,x)dx=\frac{
C_1(\alpha)K_\alpha \Delta t_1}{\alpha x_0^\alpha}.
\end{equation}
The probability to be in the vicinity of some $x\le x_0$ after $\Delta t_1$
is $f_\alpha(\Delta t_1,x)dx$. Concurrently the probability to cross the
boundary in time $\Delta t_2$ after the original time interval $\Delta t_1$ if
a walker landed at $x$ is $\int_{x_0-x}^\infty f_\alpha(\Delta t_2,x)dx$. Hence
the probability to cross the boundary within $(\Delta t_1,\Delta t_2+\Delta t_1)$
becomes
\begin{equation}
p_2=\int_{-\infty}^{x_0}dx_1\int_{x_0-x_1}^\infty f_\alpha(\Delta t_1,x_1)f_\alpha(
\Delta t_2,x_2)dx_2.
\end{equation}
For simplicity we assume that $\Delta t_1=\Delta t_2$. Then,
\begin{eqnarray}
\nonumber
p_2&=&\int_{-\infty}^\infty dx_1\int_{x_0-x_1}^\infty f_\alpha(\Delta t_1,x_1)
f_\alpha(\Delta t_1,x_2)dx_2\\\nonumber
&&-\int_{x_0}^\infty dx_1\int_{x_0-x_1}^\infty f_\alpha(\Delta t_1,x_1)f_\alpha(
\Delta t_1,x_2)dx_2\\\nonumber
&=&\int_{-\infty}^\infty dx_1\int_{x_0}^\infty f_\alpha(\Delta t_1,x_1)f_\alpha(
\Delta t_1,x_2)dx_2\\\nonumber
&&+\int_{-\infty}^\infty dx_1\int_{x_0-x_1}^{x_0} f_\alpha(\Delta t_1,x_1)f_\alpha(
\Delta t_1,x_2)dx_2\\\nonumber
&&-\int_{x_0}^\infty dx_1\int_{x_0}^\infty f_\alpha(\Delta t_1,x_1)f_\alpha(\Delta
t_1,x_2)dx_2\\\nonumber
&&-\int_{x_0}^\infty dx_1\int_{x_0-x_1}^{x_0} f_\alpha(\Delta t_1,x_1)f_\alpha(
\Delta t_1,x_2)dx_2\\
&=&p_1-p_1^2+\int_{-\infty}^{x_0} dx_1\int_{x_0-x_1}^{x_0} f_\alpha(\Delta t_1,x_1)
f_\alpha(\Delta t_1,x_2)dx_2.
\end{eqnarray}
The latter integral can be computed analytically as follows,
\begin{eqnarray}
\nonumber
\int_{-\infty}^{x_0}&dx_1&\int_{x_0-x_1}^{x_0} f_\alpha(\Delta t_1,x_1)f_\alpha(
\Delta t_1,x_2)dx_2\\
\nonumber
&=&\left(C_1(\alpha)K_\alpha \Delta t_1\right)^2\int_{-\infty}^{x_0} \frac{dx_1}{
|x_1|^{1+\alpha}}\int_{x_0-x_1}^{x_0}\frac{dx_2}{|x_2|^{1+\alpha}}\\
\nonumber
&=&\frac{\left(C_1(\alpha)K_\alpha \Delta t_1\right)^2}{x_0^{2\alpha}}\int_{-\infty}^1
\frac{dx'_1}{|x'_1|^{1+\alpha}}\int_{1-x'_1}^{1}\frac{dx'_2}{x_2^{'1+\alpha}}\\
&=&\frac{\left(C_1(\alpha)K_\alpha \Delta t_1\right)^2}{\alpha x_0^{2\alpha}}\int_{
-\infty}^1\frac{dx'_1}{|x'_1|^{1+\alpha}}\left(\frac{1}{(1-x'_1)^\alpha}-1\right).
\end{eqnarray}
In order to get an analytical result for the last integral we will split it into two
parts from $-\infty$ to 0 and from 0 to 1. The first part without the prefactor is
\begin{eqnarray}
\nonumber
\int_{-\infty}^0&\frac{dx'_1}{|x'_1|^{1+\alpha}}&\left(\frac{1}{(1-x'_1)^\alpha}-1
\right)\\
\nonumber
&=&\int_0^{\infty}\frac{dy}{y^{1+\alpha}}\left(\frac{1-(1+y)^\alpha}{(1+y)^\alpha}
\right)\\
\nonumber
&=&\int_0^\infty z^{\alpha-1}\left(\frac{1}{\left(1+1/z\right)^\alpha}-1\right)dz\\
&=&\int_0^\infty z^{2\alpha-1}\left((z+1)^{-\alpha}-z^{-\alpha}\right)dz
=B(2\alpha,-\alpha),
\end{eqnarray}
where $B(x,y)$ stands for the beta function and the last equality follows from formula
2.2.12.5 in \cite{Prudnikov} and is valid for $0<\alpha<1$. The second part produces
\begin{eqnarray}
\nonumber
\fl\int_0^1&\frac{dx'_1}{x_1^{'1+\alpha}}&\left(\frac{1}{(1-x'_1)^\alpha}-1\right)\\
\nonumber
\fl&=&\int_0^1\frac{dy}{(1-y)^{1+\alpha}}\left(\frac{1}{y^\alpha}-1\right)\\
\nonumber
\fl&=&\lim_{z\rightarrow1}\int_0^1\frac{dy}{(1-zy)^{1+\alpha}}\left(\frac{1}{y^\alpha}
-1\right)\\
\nonumber
\fl&=&\lim_{z\rightarrow1}\left\{\int_0^1y^{-\alpha+1-1}(1-y)^{2-\alpha-(1-\alpha)-1}
(1-zy)^{-1-\alpha}dy-\frac{1}{\alpha z}(1-zu)^{-\alpha}\bigg\vert_{u=0}^{u=1}
\right\}\\
\fl&=&\lim_{z\rightarrow1}\left\{\frac{\Gamma(1-\alpha)}{\Gamma(2-\alpha)}{}_2 F_1
\left(1+\alpha,1-\alpha,2-\alpha;z\right)-\frac{1}{\alpha z}\frac{1}{(1-z)^\alpha}
+\frac{1}{\alpha z}\right\},
\end{eqnarray}
where ${}_2 F_1\left(1+\alpha,1-\alpha,2-\alpha;z\right)$ is the hypergeometric
function obtained through the integral definition which is valid in this case for
$\alpha<1$. This hypergeometric function can be transformed according to 15.3.6
from \cite{abramowitz} as follows
\begin{eqnarray}
\nonumber
\fl{}_2 F_1\left(1+\alpha,1-\alpha,2-\alpha;z\right)&=&\frac{\Gamma(2-\alpha)\Gamma(
-\alpha)}{\Gamma(1-2\alpha)}{}_2 F_1\left(1+\alpha,1-\alpha,1+\alpha;1-z\right)\\
\fl&&\hspace{-2cm}
+(1-z)^{-\alpha}\frac{\Gamma(2-\alpha)\Gamma(\alpha)}{\Gamma(1+\alpha)\Gamma(1-
\alpha)}{}_2 F_1\left(1-2\alpha,1,1-\alpha;1-z\right).
\end{eqnarray}
For $z\to1$ the hypergeometric functions in the last expression will converge to 1.
Assembling all of the pieces we obtain the following formula,
\begin{equation}
\fl\int_0^1\frac{dx}{x^{1+\alpha}}\left(\frac{1}{(1-x)^\alpha}-1\right)=\frac{1}{
\alpha}+\frac{\Gamma(1-\alpha)\Gamma(-\alpha)}{\Gamma(1-2\alpha)}=\frac{1}{\alpha}
-\frac{4^\alpha\sqrt\pi\Gamma(1-\alpha)}{\alpha\Gamma(1/2-\alpha)}.
\end{equation}
Finally, collecting all the contributions to $p_2$ we get
\begin{eqnarray}
\nonumber
p_2&=&p_1-p_1^2+p_1^2 \left(1-\frac{4^\alpha\sqrt\pi\Gamma(1-\alpha)}{\Gamma(1/2
-\alpha)}-\frac{\Gamma(1-\alpha)\Gamma(2\alpha)}{\Gamma(\alpha)}\right)\\
\nonumber
&=&p_1+p_1^2\left(\frac{4^\alpha\sqrt\pi\alpha\Gamma(-\alpha)}{\Gamma(1/2-\alpha)}
-\frac{\Gamma(1-\alpha)\Gamma(2\alpha)}{\Gamma(\alpha)}\right)\\
&=&p_1-p_1^2\frac{4^\alpha\pi^{3/2}\left(1+2\cos(\pi\alpha)\right)}{\Gamma(1/2-
\alpha)\Gamma(\alpha)\sin(2\pi\alpha)}.
\label{secondjump}
\end{eqnarray} 
The last expression clearly shows that with decreasing $\alpha$ at $\alpha=2/3$
the probability to cross the boundary within an interval of
time ($\Delta t_1,2\Delta t_1$) gets lower than $p_1$ due to the change in
sign of the second term in (\ref{secondjump}). Thus, $\wp_{\mathrm{PF}}(t)$
decreases for $\alpha<2/3$ at short $t$ and at first grows for $\alpha>2/3$.

\section{Derivation of the average time of jump, the jump duration distribution
and its long-time expansion for L{\'e}vy walks, $1<\alpha\le2$}
\label{JumpDistributionsLW}

In order to estimate the effective diffusion coefficient of LWs one has to know
the distribution of durations of the jumps, the average jump duration as well as
its long-time behaviour. The characteristic function of a symmetric L{\'e}vy
stable process is
\begin{equation}
\phi(k)=\exp\Big(-\sigma_0^\alpha|k|^\alpha\Big).
\end{equation}
The distribution of the jump lengths can be expressed in terms of Fox $H$-functions
as \cite{report,Mathai}
\begin{eqnarray}
\nonumber
f_\alpha(x)=\int_0^\infty\frac{dk}{\pi}\cos(kx)e^{-\sigma_0^\alpha|k|^
\alpha}=\frac{1}{\sqrt\pi x}H^{11}_{12}\left[\frac{x^\alpha}{2^\alpha\sigma_0^
\alpha}\left|\begin{array}{l}\left(1,1\right)\\\left(\frac{1}{2},\frac{\alpha}{2}
\right),\left(1,\frac{\alpha}{2}\right)\end{array}\right.\right].
\end{eqnarray} 
The jump duration is coupled to the jump length distribution, $\tau(x)=\left|\frac{x}{
v_0}\right|$. Then the relocation time distribution reads
\begin{eqnarray}
\nonumber
\psi(\tau)&=&\int_{-\infty}^\infty dx\delta(\tau-\tau(x))f_\alpha(x)=\int_{-\infty}^
\infty dx\delta\left(\tau-\left|\frac{x}{v_0}\right|\right)f_\alpha(x)\\
\nonumber
&=&\int_{-\infty}^0 dx\delta\left(\tau+\frac{x}{v_0}\right)f_\alpha(x)+\int_0^\infty
dx\delta\left(\tau-\frac{x}{v_0}\right)f_\alpha(x)\\
&=&2\int_0^\infty dx\delta\left(\frac{x}{v_0}-\tau\right)f_\alpha(x)=2v_0f_\alpha(
v_0\tau).
\label{taudistr}
\end{eqnarray}
Hence, $\psi(\tau)$ yields in the form
\begin{equation}
\psi(\tau)=\frac{2}{\sqrt\pi\tau}H^{11}_{12}\left[\frac{v_0^\alpha}{2^\alpha\sigma_0
^\alpha}\tau^\alpha\left|\begin{array}{l}\left(1,1\right)\\\left(\frac{1}{2},\frac{
\alpha}{2}\right),\left(1,\frac{\alpha}{2}\right)\end{array}\right.\right].
\end{equation}
One can easily check that the latter function is properly normalised, $\int_0^\infty
\psi(\tau)d\tau=1$. The average jump duration can be computed for $\alpha>1$
(otherwise it diverges) as
\begin{eqnarray}
\nonumber
\left<\tau\right>&=&\int_0^\infty d\tau\tau\psi(\tau)=\frac{2}{\sqrt\pi}\int_0
^\infty d\tau H^{11}_{12}\left[\frac{v_0^\alpha}{2^\alpha\sigma_0^\alpha}\tau
^\alpha\left|\begin{array}{l}\left(1,1\right)\\\left(\frac{1}{2},\frac{\alpha}{2}
\right),\left(1,\frac{\alpha}{2}\right)\end{array}\right.\right]\\
\nonumber
&=&\frac{2}{\alpha\sqrt\pi}\int_0^\infty dt t^{1/\alpha-1}H^{11}_{12}\left[\frac{
v_0^\alpha}{2^\alpha\sigma_0^\alpha}t\left|\begin{array}{l}\left(1,1\right)\\
\left(\frac{1}{2},\frac{\alpha}{2}\right),\left(1,\frac{\alpha}{2}\right)
\end{array}\right.\right]\\
\nonumber
&=&\frac{2}{\alpha\sqrt\pi}\left(\frac{2\sigma}{v_0}\right)^{\alpha\frac{1}{
\alpha}}\Gamma\left[\begin{array}{l}\frac{1}{2}+\frac{\alpha/2}{\alpha},1-1-1/
\alpha\\1-1-\frac{\alpha/2}{\alpha}\end{array}\right]\\
&=&\frac{4}{\alpha\sqrt\pi}\frac{\sigma_0}{v_0}\frac{\Gamma\left(-\frac{1}{\alpha}
\right)}{\Gamma\left(-\frac{1}{2}\right)}=\frac{2\sigma_0}{\pi v_0}\Gamma\left(1-
\frac{1}{\alpha}\right),
\label{hsham}
\end{eqnarray}
using the properties of the $H$-function \cite{Mathai}. Our result exactly coincides
with equation (54) in \cite{BartokEli2017}, where it was derived as an approximate
formula. We show here that it is actually an exact expression (the factor $v_0$
missing in the corresponding equation of \cite{BartokEli2017} is just a misprint
there). From the distribution we obtain the exact form in Laplace space and expand
it at small $s$ corresponding to long times. The Laplace transform of equation
(\ref{taudistr}) reads \cite{Mathai}
\begin{eqnarray}
\nonumber
\mathscr{L}(\psi(\tau))&=&\tilde\psi(s)=\mathscr{L}\left(\frac{2}{\sqrt\pi\tau}
H^{11}_{12}\left[\frac{v_0^\alpha}{2^\alpha\sigma_0^\alpha}\tau^\alpha\left|
\begin{array}{l}\left(1,1\right)\\\left(\frac{1}{2},\frac{\alpha}{2}\right),\left(1,
\frac{\alpha}{2}\right)\end{array}\right.\right]\right)\\
\nonumber
&=&\frac{2}{\sqrt\pi}H^{12}_{22}\left[\frac{v_0^\alpha}{2^\alpha\sigma_0^\alpha}s^{
-\alpha}\left|\begin{array}{l}\left(1,\alpha\right),\left(1,1\right)\\\left(\frac{
1}{2},\frac{\alpha}{2}\right),\left(1,\frac{\alpha}{2}\right)\end{array}\right.\right]\\
&=&\frac{2}{\sqrt\pi}H^{21}_{22}\left[\frac{2^\alpha\sigma_0^\alpha}{v_0^\alpha}s^{
\alpha}\left|\begin{array}{l}\left(\frac{1}{2},\frac{\alpha}{2}\right),\left(0,
\frac{\alpha}{2}\right)\\\left(0,\alpha\right),\left(0,1\right)\end{array}\right.
\right].
\label{laplacetdistr}
\end{eqnarray}
The expansion of the latter $H$-function for $s\rightarrow0$ has a ratio of nominally
infinite values due to the presence of zeros among the coefficients. In order to solve
this problem we introduce the infinitesimal parameter $\varepsilon$ and treat the
$H$-function as a limit of another $H$-function with non-zero coefficients
$\varepsilon$, namely,
\begin{eqnarray}
\fl H^{21}_{22}\left[\frac{2^\alpha\sigma_0^\alpha}{v_0^\alpha}s^{\alpha}\left|
\begin{array}{l}\left(\frac{1}{2},\frac{\alpha}{2}\right),\left(0,\frac{\alpha}{2}
\right)\\
\left(0,\alpha\right),\left(0,1\right)\end{array}\right.\right]=\lim_{\varepsilon
\to0}H^{21}_{22}\left[\frac{2^\alpha\sigma_0^\alpha}{v_0^\alpha}s^{\alpha}\left|
\begin{array}{l}\left(\frac{1}{2},\frac{\alpha}{2}\right),\left(\varepsilon,
\frac{\alpha}{2}\right)\\\left(\varepsilon,\alpha\right),\left(\varepsilon,1\right)
\end{array}\right.\right].
\end{eqnarray}
The expansion of the latter $H$-function reads (\cite{Prudnikov,Mathai}),
\begin{eqnarray}
\nonumber
\fl H^{21}_{22}\left[z\left|\begin{array}{l}\left(\frac{1}{2},\frac{\alpha}{2}
\right),\left(\varepsilon,\frac{\alpha}{2}\right)\\\left(\varepsilon,\alpha\right)
,\left(\varepsilon,1\right)\end{array}\right.\right]&=&\sum_{k=0}^\infty\frac{2
\Gamma\left(\varepsilon-\frac{1}{\alpha}(\varepsilon+k)\right)\Gamma\left(\frac{1}{
2}+\frac{1}{2}(\varepsilon+k)\right)}{\sqrt\pi\Gamma\left(\varepsilon-\frac{1}{2}(
\varepsilon+k)\right)}\frac{(-1)^kz^{(\varepsilon+k)/\alpha}}{k!\alpha}\\
&&\hspace*{-2cm}
+\sum_{k=0}^\infty\frac{2\Gamma\left(\varepsilon-\alpha(\varepsilon+k)\right)\Gamma
\left(\frac{1}{2}+\frac{1}{2}(\varepsilon+k)\right)}{\sqrt\pi\Gamma\left(\varepsilon
-\frac{1}{2}(\varepsilon+k)\right)}\frac{(-1)^kz^{\varepsilon+k}}{k!},
\end{eqnarray}
where $z=2^\alpha\sigma_0^\alpha s^{\alpha}/(v_0^\alpha)$. Using the known behaviour
of the Gamma function for small arguments, $\Gamma(\varepsilon)\sim1/\varepsilon$,
we obtain the following expansion,
\begin{equation}
\psi(s)\sim1-\left<\tau\right>s+As^\alpha+\ldots,
\end{equation}
where 1 has contributions from zeroth orders of both sums, $\left<\tau\right>$ is
given by expression (\ref{hsham}), and A reads
\begin{eqnarray}
A=-\frac{2^{\alpha+1}\sigma_0^\alpha}{\sqrt\pi v_0^\alpha}\frac{\Gamma(-\alpha)
\Gamma\left(1/2+\alpha/2\right)}{\Gamma\left(-\alpha/2\right)}=\frac{\sigma_0^{
\alpha}}{v_0^\alpha|\cos(\pi\alpha/2)|}.
\end{eqnarray}
This form can be shown to be equivalent to the corresponding expression in
\cite{BartokEli2017}.

\section{Derivation of the effective long-time diffusion coefficient of LWs,
$1<\alpha\le2$}
\label{LWdiffcoefDerivation}

Let us start from equation (29) in \cite{RevModPhys2015},
\begin{equation}
P(k,s)=\frac{\left[\tilde\Psi(s+ikv)+\tilde\Psi(s-ikv)\right]P_0(k)}{2-\left[\psi(
s+ikv)+\psi(s-ikv)\right]},
\end{equation} 
where $\tilde\Psi(s)$ is the Laplace transform of the survival probability and can
be expressed through $\psi(s)$ as $\tilde\Psi(s)=[1-\psi(s)]/s$, and for long times
(small $s$) we have $\psi(s)\sim1-\left<\tau\right>s+As^\alpha$. Hence,
\begin{equation}
P(k,s)\sim\frac{P_0(k)\left[2\left<\tau\right>-A\left((s+ikv)^{\alpha-1}+(s-ikv)^{
\alpha-1}\right)\right]}{2s\left<\tau\right>-A\left((s+ikv)^{\alpha}+(s-ikv)^{\alpha}
\right)}.
\end{equation}
The expressions in the round brackets can be expanded,
\begin{equation}
(s\pm ikv)^\alpha\sim e^{\pm\frac{i\pi\alpha}{2}}|k|^\alpha v_0^\alpha\left(1\pm
\frac{s\alpha}{ikv}\right).
\end{equation}
The last expansion assumes the condition $s\ll|k|v_0\ll1$. After neglecting higher
powers in $s$ one gets
\begin{equation}
P(k,s)=\frac{P_0(k)}{s-A|k|^\alpha v_0^\alpha\cos\left(\frac{\pi\alpha}{2}\right)
/\left<\tau\right>}.
\end{equation}
Alternatively,
\begin{equation}
P(k,s)=\frac{P_0(k)}{s+K^{LW}_\alpha|k|^\alpha},
\end{equation}
where
\begin{eqnarray}
\nonumber
K_\alpha^{\mathrm{LW}}&=&\frac{Av_0^\alpha\left|\cos\left(\frac{\pi\alpha}{2}\right)
\right|}{\left<\tau\right>}=\frac{2\sigma_0^\alpha\sin\left(\frac{\pi\alpha}{2}\right)
\Gamma(-\alpha)\Gamma(1+\alpha)\left|\cos\left(\frac{\pi\alpha}{2}\right)\right|}{\pi
\left<\tau\right>}\\
\nonumber
&=&\frac{\sigma_0^{\alpha-1}\sin\left(\frac{\pi\alpha}{2}\right)\Gamma(-\alpha)\Gamma(
1+\alpha)\left|\cos\left(\frac{\pi\alpha}{2}\right)\right|v_0}{\Gamma\left(1-\frac{
1}{\alpha}\right)}\\
&=&\frac{\sigma_0^{\alpha}}{\langle\tau\rangle},
\label{keff}
\end{eqnarray}
where $\langle\tau\rangle$ is given by equation (\ref{hsham}). The last result is
exactly the intuitive definition of the diffusivity from a continuous time random
walk perspective. The second line in equation (\ref{keff}) corrects the formula (55)
in \cite{BartokEli2017}.

\section{Estimation of the jump in the first-passage time PDF of LWs}
\label{JumpPWderivation}

The jump in the first-passage time PDF of LWs corresponds to the probability stored
in the ballistically moving front peak of the position PDF \cite{RevModPhys2015}.
The density of particles in these peaks reads (equation (32) in \cite{RevModPhys2015})
\begin{equation}
G_{\mathrm{front}}(x,t)=\frac{1}{2}\Psi(t)\Big(\delta(x-vt)+\delta(x+vt)\Big).
\end{equation} 
From the equations above we can compute the survival probability as a function of
time. In Laplace space $\tilde\Psi(s)=[1-\psi(s)]/s$. From expression
(\ref{laplacetdistr}),
\begin{eqnarray}
\frac{\tilde\psi(s)}{s}=\frac{2}{\sqrt\pi s}H^{21}_{22}\left[\frac{2^\alpha\sigma_0
^\alpha}{v_0^\alpha}s^{\alpha}\left.\begin{array}{l}\left(\frac{1}{2},\frac{\alpha}{2}
\right),\left(0,\frac{\alpha}{2}\right)\\\left(0,\alpha\right),\left(0,1\right)
\end{array}\right.\right].
\end{eqnarray} 
Its inverse Laplace transform reads \cite{Mathai},
\begin{eqnarray}
\nonumber
\mathscr{L}^{-1}\left(\frac{\tilde\psi(s)}{s}\right)&=&\frac{2}{\sqrt\pi}H^{21}_{32}
\left[\frac{2^\alpha\sigma_0^\alpha}{v_0^\alpha}t^{-\alpha}\left|\begin{array}{ll}
\left(\frac{1}{2},\frac{\alpha}{2}\right),\left(0,\frac{\alpha}{2}\right)&(1,\alpha)\\
\left(0,\alpha\right),\left(0,1\right)\end{array}\right.\right]\\
&=&\frac{2}{\sqrt\pi}H^{12}_{23}\left[\frac{v_0^\alpha}{2^\alpha\sigma_0^\alpha}t^{
\alpha}\left|\begin{array}{l}\left(1,\alpha\right),\left(1,1\right)\\\left(\frac{1}{2},
\frac{\alpha}{2}\right),\left(1,\frac{\alpha}{2}\right),(0,\alpha)\end{array}\right.
\right].
\end{eqnarray}
The survival probability as a function of $t$ is then
\begin{eqnarray}
\Psi(t)=1-\frac{2}{\sqrt\pi}H^{12}_{23}\left[\frac{v_0^\alpha}{2^\alpha\sigma_0^
\alpha}t^{\alpha}\left|\begin{array}{l}\left(1,\alpha\right),\left(1,1\right)\\
\left(\frac{1}{2},\frac{\alpha}{2}\right),\left(1,\frac{\alpha}{2}\right),(0,\alpha)
\end{array}\right.\right].
\end{eqnarray}
This is an exact analytical result. The value of this $H$-function can be found from
an expansion in $t$ \cite{Prudnikov},
\begin{eqnarray}
\nonumber
&&H^{12}_{23}\left[\frac{v_0^\alpha}{2^\alpha\sigma_0^\alpha}t^{\alpha}\left|
\begin{array}{l}\left(1,\alpha\right),\left(1,1\right)\\
\left(\frac{1}{2},\frac{\alpha}{2}\right),\left(1,\frac{\alpha}{2}\right),(0,\alpha)
\end{array}\right.\right]\\
&&\hspace*{2.2cm}
=\frac{2}{\alpha}\sum_{k=0}^\infty\frac{\Gamma(2k+1)\Gamma\left(\frac{2k+1}{\alpha}\right)}{\Gamma\left(k+\frac{1}{2}\right)\Gamma(2+2k)}\left(\frac{v_0t}{2\sigma_0}\right)^{1+2k}\frac{(-1)^k}{k!}
\label{Gfrontseries}
\end{eqnarray}
The last two equations allow one to compute the value of $G_{\mathrm{front}}$.
Unless the argument of the $H$-function is too big the series (\ref{Gfrontseries})
converges quite fast.

\section{Derivation of the long-time limit of the first-hitting time PDF of LFs,
$1<\alpha\le2$}
\label{LongTimeAF}

We start from equation (\ref{twointsolution}). The integral in the numerator can be
computed analytically and reads
\begin{equation}
\int_{-\infty}^{\infty}dk\frac{1}{s+K_\alpha\left\vert k \right\vert^{\alpha}}=
\frac{2\pi}{\alpha\sin(\pi/\alpha)}\frac{s^{1/\alpha-1}}{K_\alpha^{1/\alpha}}.
\end{equation}
Hence equation (\ref{twointsolution}) can be rewritten as
\begin{equation}
\wp_{\mathrm{HF}}(s)=\frac{\alpha\sin(\pi/\alpha)K_\alpha^{1/\alpha}}{\pi}\int_0^{
\infty}dk\frac{s^{1-1/\alpha}}{s+K_\alpha|k|^{\alpha}}\cos(kx_0).
\end{equation}
From the relation (1.80) in \cite{Podlubny} we know that
\begin{equation}
t^{\beta-1}E_{\alpha,\beta}(-at^\alpha)\div\frac{s^{\alpha-\beta}}{s^{\alpha}+a},
\end{equation}
where $\div$ stands as a notation for the Laplace transform pair and $E_{\alpha,
\beta}(t)$ is the two-parameter Mittag-Leffler function. Hence, we get equation
(\ref{LFFApdftime}),
\begin{eqnarray}
\wp_{\mathrm{HF}}(t)=\frac{\alpha\sin\left(\pi/\alpha\right)K_\alpha^{1/\alpha}}{
\pi}t^{1/\alpha-1}\int_0^\infty dk\cos(kx_0)E_{1,1/\alpha}(-K_\alpha k^{\alpha}t).
\label{FAtimespacesolution}
\end{eqnarray}
This equation clearly shows that for $\alpha=1$ the PDF of first-hitting the target
is exactly zero, $\wp_{\mathrm{HF}}=0$ (the prefactor is equal to zero in this case).
Now we consider the long-time limit $t\rightarrow\infty$. Rewriting the previous
equation as
\begin{eqnarray}
\nonumber
\fl\wp_{\mathrm{HF}}(t)&=&\frac{\alpha\sin\left(\pi/\alpha\right)K_\alpha^{1/\alpha}}{
\pi}t^{1/\alpha-1}\int_0^\infty dkE_{1,1/\alpha}(-K_\alpha k^{\alpha}t)\\
\fl&&-\frac{\alpha\sin\left(\pi/\alpha\right)K_\alpha^{1/\alpha}}{\pi}t^{1/\alpha-1}
\int_0^\infty dk(1-\cos(kx_0))E_{1,1/\alpha}(-K_\alpha k^{\alpha}t).
\end{eqnarray}
Consider separately the first and the second contribution. The first integral is zero,
\begin{eqnarray}
\nonumber
\int_0^\infty dkE_{1,1/\alpha}(-K_\alpha k^{\alpha}t)&\sim&(K_\alpha t)^{-1/\alpha}
\int_0^{\infty}dk'E_{1,1/\alpha}(-k'^{\alpha})\\
\nonumber
&\sim&\frac{1}{\alpha}(K_\alpha t)^{-1/\alpha}\int_0^\infty y^{1/\alpha-1}E_{1,1/
\alpha}(-y)dy\\
\nonumber
&\sim&\lim_{z\to\infty}\int_0^zy^{1/\alpha-1}E_{1,1/\alpha}(-y)dy\\
\nonumber
&\sim&\lim_{z\rightarrow\infty}z^{1/\alpha}E_{1,1+1/\alpha}(-z)\\
&\sim&\lim_{z\rightarrow\infty}z^{1/\alpha}\frac{1}{z}=0,
\end{eqnarray}
where we used formula (1.99) in \cite{Podlubny}. In the second integral the
Mittag-Leffler function can be substituted by its large argument limit, $E_{1,1/
\alpha}(-K_\alpha k^{\alpha}t)\to-1/[K_\alpha k^\alpha t\Gamma(1/\alpha-1)]$,
because at small values of $k$ the expression $(1-\cos(kx_0))$ disappears. Then,
\begin{eqnarray}
\nonumber
\wp_{\mathrm{HF}}(t)&=&-\frac{\alpha\sin\left(\frac{\pi}{\alpha}\right)K_\alpha^{
1/\alpha-1}}{\pi\Gamma(1/\alpha-1)t^{1/\alpha-2}}\int_0^\infty\frac{1-\cos(kx_0)}{
k^\alpha}dk\\
\nonumber
&=&-\frac{\alpha\sin\left(\frac{\pi}{\alpha}\right)}{\pi\Gamma(1/\alpha-1)}\frac{
\Gamma(2-\alpha)\sin\left(\frac{\pi\alpha}{2}\right)}{\alpha-1}x_0^{\alpha-1}K_
\alpha^{1/\alpha-1}t^{1/\alpha-2}\\
&=&\frac{\sin\left(\frac{\pi}{\alpha}\right)\Gamma(2-\alpha)\sin\left(\frac{\pi
\alpha}{2}\right)}{\pi\Gamma(1/\alpha)}x_0^{\alpha-1}K_\alpha^{1/\alpha-1}t^{1/
\alpha-2},
\label{longAFanalyt}
\end{eqnarray}
which exactly coincides with result (A.9) in \cite{JPA16} and is equivalent to the
corresponding expression in \cite{JPA2003}.

\section{Derivation of the short-time limit of the first-hitting time PDF of LFs}
\label{ShortTimeAF}

Here we compute the power-law of the first-hitting time PDF of LFs in the limit
of short times. It is again convenient to start from equation (\ref{LFFApdftime}).
We change the variable as $y=k(K_\alpha t)^{1/\alpha}$. Hence,
\begin{equation}
\wp_{\mathrm{HF}}(t)=\frac{\alpha\sin\left(\pi/\alpha\right)}{\pi}t^{-1}\int_0
^\infty dy\cos\left(\frac{yx_0}{(K_\alpha t)^{1/\alpha}}\right)E_{1,1/\alpha}(
-y^\alpha).
\end{equation}
Introducing the notation $\lambda=x_0/(K_\alpha t)^{1/\alpha}$ and using the definition
of the Mittag-Leffler function,
\begin{eqnarray}
\nonumber
\fl\wp_{\mathrm{HF}}(t)&=&\frac{\alpha\sin\left(\pi/\alpha\right)}{\pi}t^{-1}\int_0
^\infty dy\cos\left(\lambda y\right)\sum_{n=0}^{\infty}\frac{(-y^\alpha)^n}{\Gamma(
n+1/\alpha)}\\
\fl&=&\frac{\alpha\sin\left(\pi/\alpha\right)}{\pi}t^{-1}\sum_{n=0}^{\infty}\int_0
^\infty dy\cos\left(\lambda y\right)\frac{(-y^\alpha)^n}{\Gamma(n+1/\alpha)}=\sum_{
n=0}^\infty I_n(t),
\label{expansion1}
\end{eqnarray}
where $I_n(t)$ is the $n$-th term of the expansion of $\wp_{\mathrm{HF}}(t)$ in a
power series. With the help of equation (II.25) for improper integrals in
\cite{Malakhov},
\begin{equation}
\int_0^\infty\tau^ke^{-i\omega\tau}d\tau=\frac{\Gamma(k+1)}{\omega^{k+1}}e^{-\frac{
1}{2}(k+1)\pi i}
\end{equation}
we compute the terms one by one. For convenience we use the shorthand notation
$C(\alpha)=\frac{\alpha\sin\left(\pi/\alpha\right)}{\pi}$. Then,
\begin{eqnarray}
\nonumber
I_0(t)&=&C(\alpha)t^{-1}\frac{1}{\Gamma(1/\alpha)}\int_0^\infty dy\cos(\lambda y)
=0,\\
\nonumber
I_1(t)&=&C(\alpha)t^{-1}\frac{-1}{\Gamma(1+1/\alpha)}\frac{\Gamma(\alpha+1)}{
\lambda^{\alpha+1}}\mathrm{Re}\left[(-i)^{\alpha+1}\right]\\
\nonumber
&=&C(\alpha)t^{-1}\frac{-1}{\Gamma(1+1/\alpha)}\frac{\Gamma(\alpha+1)}{\lambda^{
\alpha+1}}\mathrm{Re}\exp\left(-\frac{i\pi}{2}(1+\alpha)\right)\\
\nonumber
&=&-C(\alpha)t^{-1}\frac{-1}{\Gamma(1+1/\alpha)}\frac{\Gamma(\alpha+1)}{\lambda^{
\alpha+1}}\sin\left(\frac{\pi\alpha}{2}\right)\\
\label{I1}
&=&C(\alpha)\frac{\Gamma(1+\alpha)\sin(\pi\alpha/2)}{\Gamma(1+1/\alpha)}\frac{
K_\alpha^{1+1/\alpha}}{x_0^{\alpha+1}}t^{1/\alpha},\\
\label{I2}
I_2(t)&=&C(\alpha)\frac{\Gamma(2\alpha+1)}{\Gamma(2+1/\alpha)(-\sin(\pi\alpha))}
\frac{K_\alpha^{2+1/\alpha}}{x_0^{2\alpha+1}}t^{1+1/\alpha},\\
\label{I3}
I_3(t)&=&C(\alpha)\frac{\Gamma(3\alpha+1)}{\Gamma(3+1/\alpha)(-\sin\left(3\pi
\alpha/2\right))}\frac{K_\alpha^{3+1/\alpha}}{x_0^{3\alpha+1}}t^{2+1/\alpha}.
\end{eqnarray}

\section{The target size selection in the simulations}
\label{targetsizeappendix}

\begin{figure}
\centering
\includegraphics[width=11.2cm]{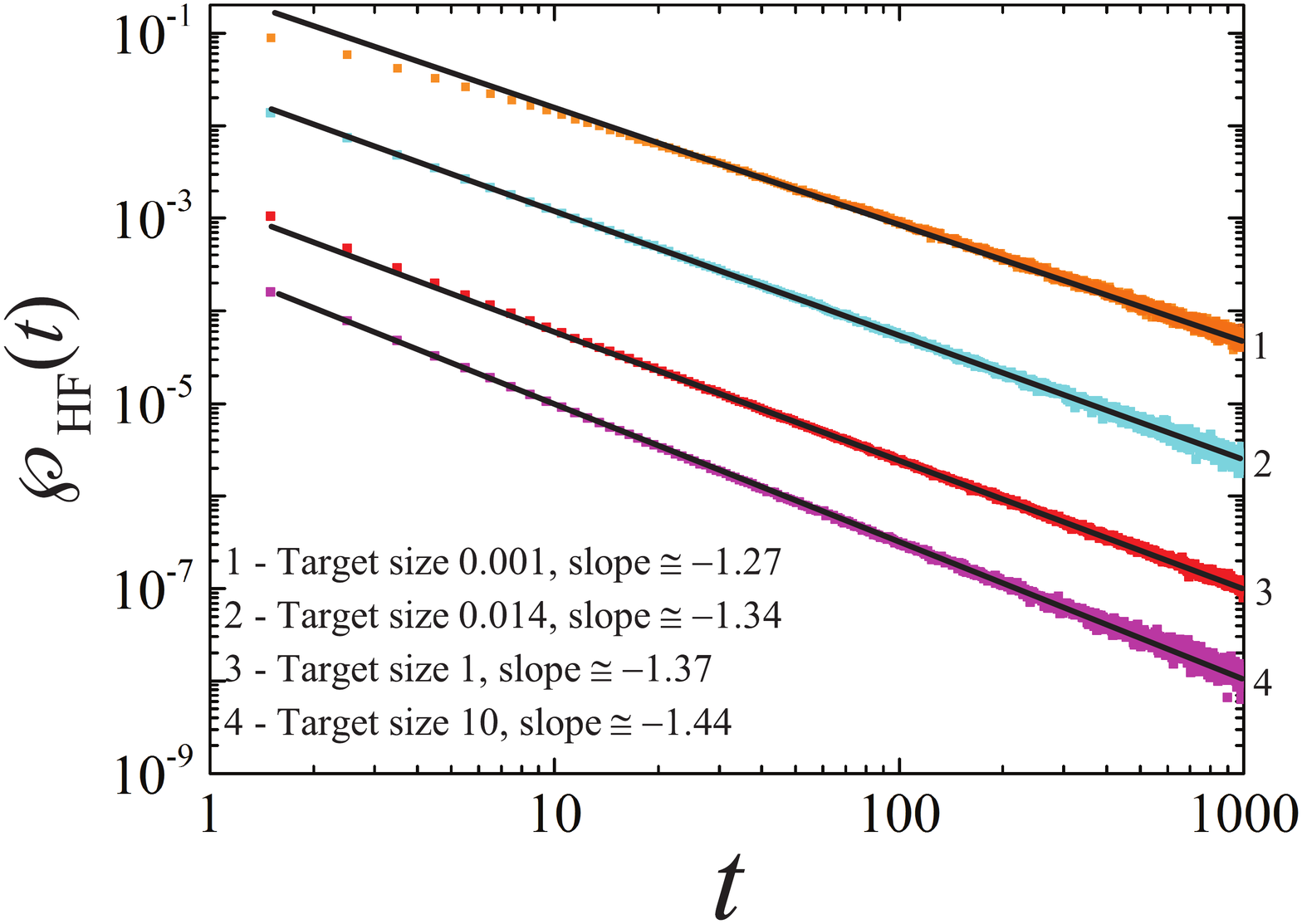}
\caption{Dependence of the long-time exponent of the PDF of first-hitting on
the target size for LFs. Note that for better visual comparison of the PDFs
their values were divided by a factor of 10 for $d=0.014$, of 100 for $d=1$,
and of 1000 for $d=\infty$. In the first three cases the target was centred
at $x=0$. For $d=10$ the target is centred at $x=-5$. Parameters: $\alpha=1.5$,
$K_\alpha=1$, $x_0=1$, and $\delta t=0.001$. The number of runs is $2\times10^6$.}
\label{TS1}
\end{figure}

\begin{figure}
\centering
\includegraphics[width=11.2cm]{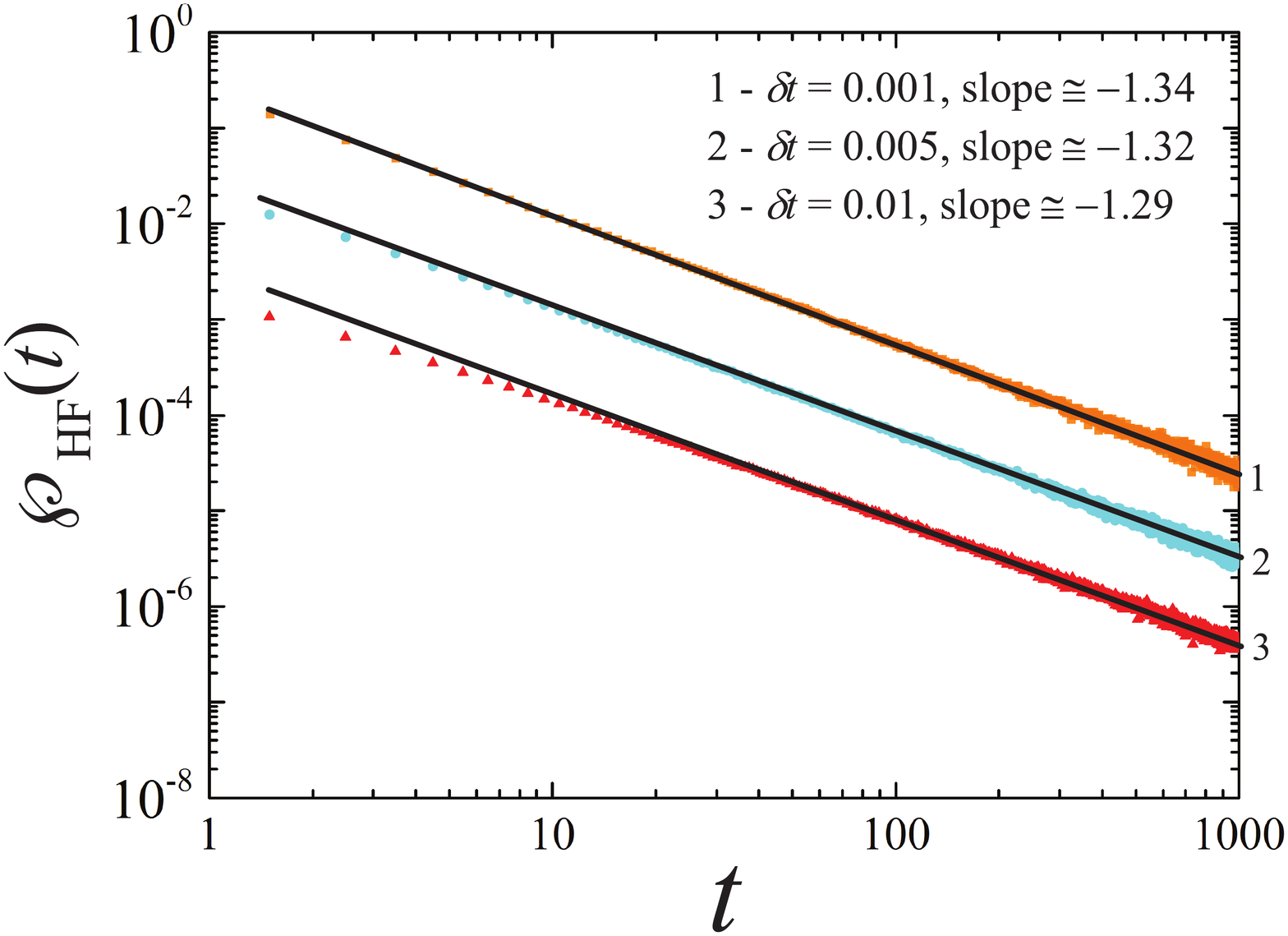}
\caption{Dependence of the long-time exponent of the PDF of first-hitting
on the time resolution in the simulations of LFs when the target size is
fixed (the best choice for the $\delta t=0.001$ was selected). Note that
for better visual comparison of the PDFs their values were divided
by a factor of 10 for $\delta t=0.005$ and of 100 for $\delta t=0.01$. Parameters:
$\alpha=1.5,K_\alpha=1$, $x_0=1$, and $d=0.014$.}
\label{TS2}\end{figure}

Our theoretical framework for LFs includes the notion of a point-like target.
However, in the simulations the target size should be non-zero if the target is to
be successfully located. Thus the target should not be too small such that it can
actually be hit successfully. At the same time it should not be too large, otherwise
this would result in the scenario of the first-passage, and thus the scaling exponent
would converge to the universal Sparre-Andersen result. One has to choose the target
size correctly. From the theory the asymptotic behaviour at long times is known to be
$\wp_{\mathrm{HF}}(t)\simeq t^{1/\alpha-2}$ \cite{JPA2003}. In figure G1 we show how
the exponent at long times depends on the target size for the time resolution $\delta
t=0.001$. For very small targets the exponent is closer to -1 (data 1 in figure
\ref{TS1}) and the Sparre-Andersen limit can be obtained when large targets are
considered (data set 4 in figure \ref{TS1} corresponds to the case in which the
starting point was distance 1 away from the boundary of the target with $d=10$). We
also checked that this automatically leads to the proper power law with scaling
exponent $1/\alpha$ at short times. Vice versa
one could use the short-time power law to get the best fit target size with the
correct long-term first-hitting PDF statistics. Note that in order to get the
exponents in figure \ref{TS1} we averaged the data in the time interval $[50,1000]$.
For any fixed $d$ and $\delta t$ the exponent decreases slightly when the frame of
averaging is moved towards longer $t$. No matter how small the target is chosen, the
proper point-like target long-time scaling will be observed at some time scale.
However, any simulation has a finite duration. Hence, it is reasonable to choose a
target size which leads to the theoretical scaling within the time cutoff limit of
the run. The set of best choices for other $\alpha$ used is listed in the main text.

Since the overshoots also depend on the time discretisation one has to adjust it as
well. In figure \ref{TS2} we show how the long time characteristics change with
$\delta t$ while keeping the target size $d=0.014$ for $\alpha=1.5$. The change of
$\delta t$ leads to deviations from the proper long-time behaviour of the first-hitting
PDF. Importantly, the number of runs does not affect the exponent obtained by
averaging over a fixed interval of time. However, the accuracy of defining of the
exponents increases with the number of runs.

\section{Shape of the L{\'e}vy walk propagator}

\begin{figure}
\centering
\includegraphics[width=11.2cm]{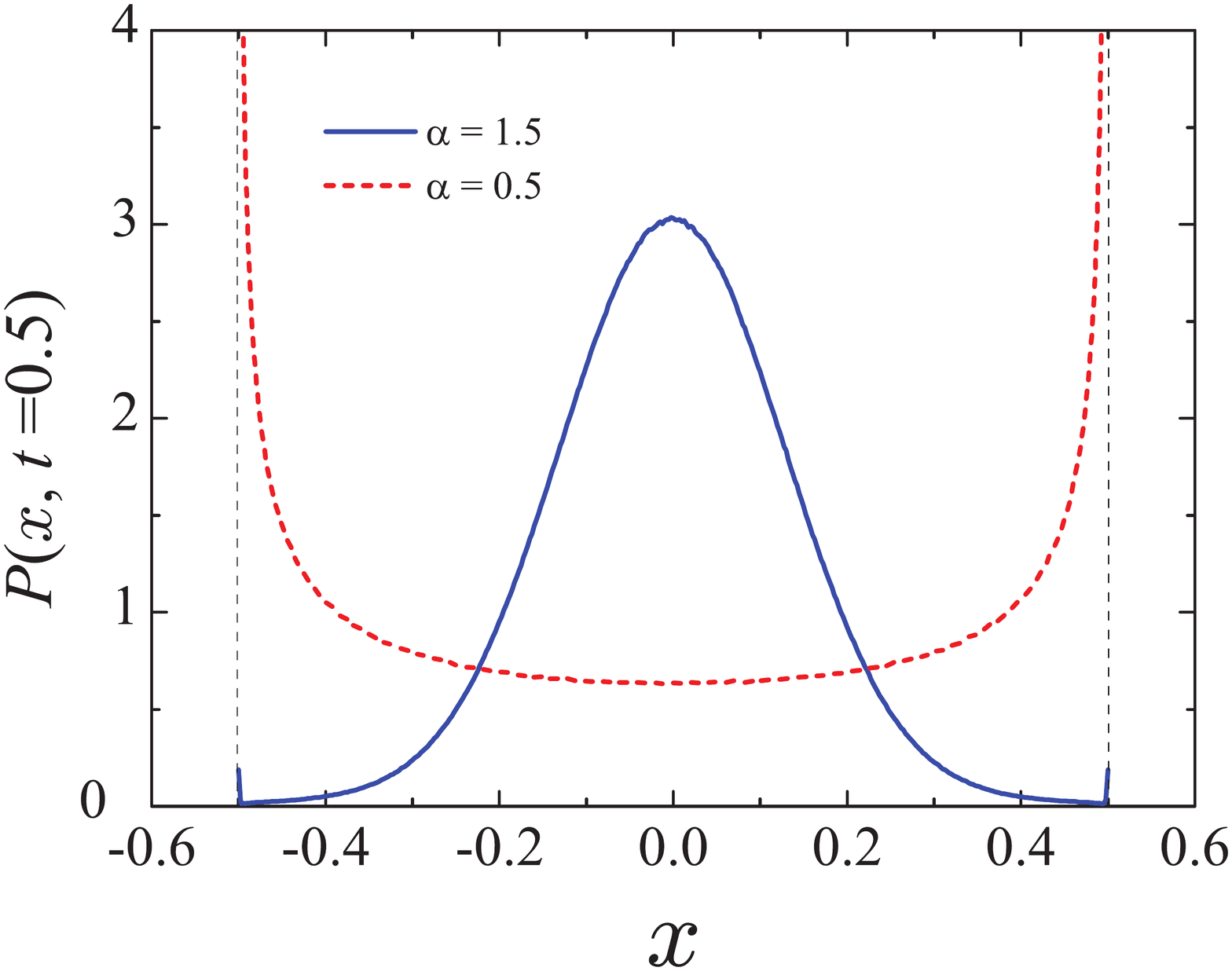}
\caption{LW PDF for $\alpha=0.5$ and $\alpha=1.5$ at $t=0.5$, for $v=1$. For
$\alpha=0.5$ the PDF is U-shaped while for $\alpha=1.5$ the PDF is bell-shaped.
In both cases the horizon of the PDF is at $|x|=0.5$, note the spikes at
$|x|=0.5$ for $\alpha=1.5$.}
\label{supp_pdf}
\end{figure}

The shape of the LW propagator discussed in section \ref{fplw} is shown in
figure \ref{supp_pdf}.

\ack
VVP acknowledges fruitful discussions with Olivier B\'enichou. RM and AVC acknowledge
financial support from Deutsche Forschungsgemeinschaft (Grant numbers ME 1535/6-1
and ME 1535/7-1). RM also acknowledges an Alexander von Humboldt Honorary Research
Scholarship from the Foundation for Polish Science. NW and GB were funded at
Potsdam and Dresden, respectively, by ONR Global NICOP grant N62909-15-1-N143 to
the CFSA, University of Warwick, and they wish to thank the PI, Sandra Chapman.
RK thanks Sabine Klapp and Holger Stark, TU Berlin, for their hospitality as a guest
scientist as well as ONR Global for financial support. He also acknowledges funding
from the London Mathematical Laboratory, where he is an External Fellow.

\end{document}